\documentclass{article}

\usepackage{iclr2024_conference,times}
\usepackage[utf8]{inputenc} 
\usepackage[T1]{fontenc}    
\usepackage{hyperref}       
\usepackage{url}            
\usepackage{booktabs}
\usepackage{multirow}
\usepackage{amsfonts}       
\usepackage{nicefrac}       
\usepackage{microtype}      
\usepackage{xcolor}         
\usepackage{wrapfig}
\usepackage{etoolbox}
\BeforeBeginEnvironment{wrapfigure}{\setlength{\intextsep}{-6pt}}

\usepackage{graphicx}
\usepackage{subcaption}
\usepackage{lipsum}
\usepackage{enumitem}

\usepackage[ruled, lined, linesnumbered, commentsnumbered, longend, noend]{algorithm2e}

\usepackage{amsmath}
\usepackage{amssymb}
\usepackage{mathtools}
\usepackage{amsthm}

\usepackage[capitalize,noabbrev]{cleveref}

\theoremstyle{plain}
\newtheorem{theorem}{Theorem}[section]

\newtheorem{lemma}[theorem]{Lemma}

\theoremstyle{definition}
\newtheorem{definition}[theorem]{Definition}
\newtheorem{assumption}[theorem]{Assumption}
\theoremstyle{remark}
\newtheorem{remark}[theorem]{Remark}
\usepackage{bm}

\DeclareMathAlphabet{\mathcal}{OMS}{cmsy}{m}{n}

\DeclareMathOperator{\EX}{\mathbb{E}}
\usepackage{dsfont}
\usepackage[textsize=tiny]{todonotes}

\newcommand\blfootnote[1]{%
  \begingroup
  \renewcommand\thefootnote{}\footnote{#1}%
  \addtocounter{footnote}{-1}%
  \endgroup
}

\usepackage{xargs}
\usepackage{xcolor}
\definecolor{BrickRed}{rgb}{0.8, 0.25, 0.33}
\ifx\commentsoff\undefined
    
\fi

\title{\textsc{FedQV}: Leveraging Quadratic Voting in Federated Learning}

\author{Tianyue Chu 
\\
IMDEA Networks Institute\\
Universidad Carlos III de Madrid\\
\And
Nikolaos Laoutaris \\
IMDEA Networks Institute \\
Madrid, Spain \\
}

\begin{document}

\maketitle

\begin{abstract}
Federated Learning (FL) permits different parties to collaboratively train a global model without disclosing their respective local labels. A crucial step of FL, that of aggregating local models to produce the global one, shares many similarities with public decision-making, and elections in particular. In that context, a major weakness of FL, namely its vulnerability to poisoning attacks, can be interpreted as a consequence of the \emph{one person one vote} (henceforth \emph{1p1v}) principle underpinning most contemporary aggregation rules. 
In this paper, we propose \textsc{FedQV}, a novel aggregation algorithm built upon the \emph{quadratic voting} scheme, recently proposed as a better alternative to \emph{1p1v}-based elections. Our theoretical analysis establishes that \textsc{FedQV} is a truthful mechanism in which bidding according to one's true valuation is a dominant strategy that achieves a convergence rate that matches those of state-of-the-art methods. Furthermore, our empirical analysis using multiple real-world datasets validates the superior performance of \textsc{FedQV} against poisoning attacks. It also shows that combining \textsc{FedQV} with unequal voting ``budgets'' according to a reputation score increases its performance benefits even further. Finally, we show that \textsc{FedQV} can be easily combined with Byzantine-robust privacy-preserving mechanisms to enhance its robustness against both poisoning and privacy attacks. 

\blfootnote{Please cite the ACM SIGMETRICS'24 version of this paper}
\end{abstract}

\section{Introduction}
Federated Learning (FL) has emerged as a promising privacy-preserving paradigm for conducting distributed collaborative model training across parties that do not want to disclose their local data.  
Agreeing on a common global model in Federated Learning shares many similarities with public decision-making and elections in particular. Indeed, the weights of local model updates of a party (client) can be seen as votes of preference that affect the global model resulting from an aggregation rule applied at the centralised server of an FL group.
\textsc{FedAvg}~\cite{mcmahan2017communication} has been the ``de facto'' aggregation rule used in FL tasks such as Google's emoji and next-word prediction for mobile device keyboards~\cite{ramaswamy2019federated,hard2018federated}. In \textsc{FedAvg} the global model is produced from a simple weighted averaging of local updates with weights that represent the amount of data that each party has used for its training. 

\paragraph{The problem}
Recent work~\cite{blanchard2017machine} has shown that \textsc{FedAvg} is vulnerable to poisoning attacks, as even a single attacker can degrade the global model by sharing faulty local updates of sufficiently large weight. Such attacks become possible because \textsc{FedAvg} treats all local data points equally. In essence, the aggregation rule, when seen at the granularity of individual training data, resembles the \emph{one person one vote (1p1v)} election rule of modern democratic elections. In this context, the server distributes votes (weights) to a party in accordance with the amount of its training data, which may be regarded as its population. This, however, may confer an unjust advantage to malicious parties with large training datasets.


\paragraph{Our approach} To address this issue, we propose a novel aggregation rule inspired by elections based on \emph{Quadratic Voting} ~\cite{lalley2018quadratic} (henceforth QV). 
In QV, each party is given a voting budget that can be spent on different rounds of voting. 
Within a particular vote, an individual has to decide the number of "credit voices" to commit, whose square root is what impacts the corresponding outcome of the vote. 
QV has been proposed as a means
to break out from the tyranny-of-the-majority vs. subsidising-the-minority 
dilemma of election systems~\cite{posner2015voting}. Its formal analysis~\cite{weyl2017robustness} under a game theoretic price-taking model, has shown that QV outperforms \emph{1p1v} in terms of efficiency and robustness. Importantly, it has the unique capacity to deter collusion attacks by effectively taxing extreme behaviours.

\begin{wrapfigure}[21]{r}[0pt]{0.3\textwidth}
  \centering
  \includegraphics[width=0.3\textwidth]{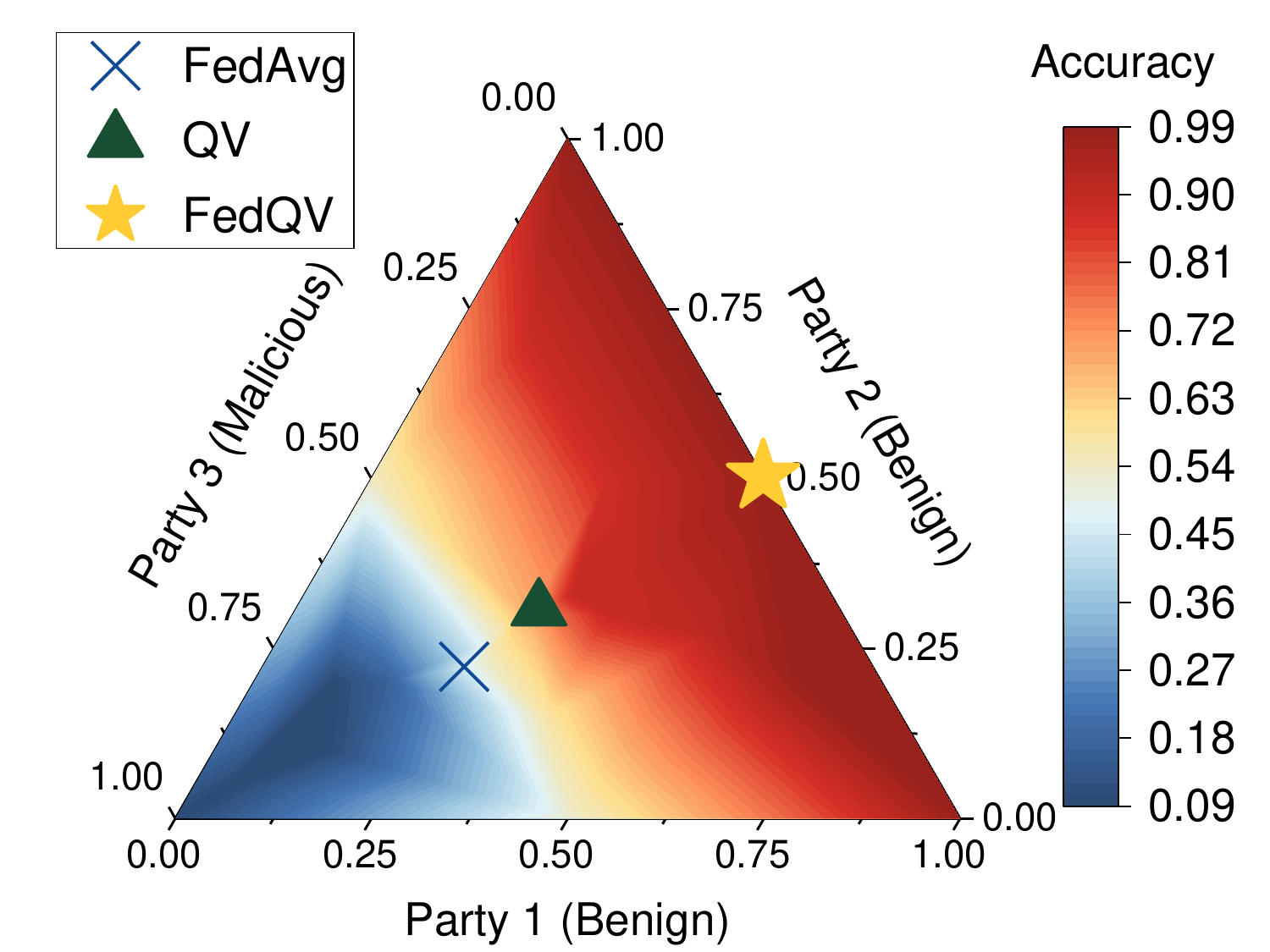} 
  \caption{Global model weights (position within the triangle) and corresponding test accuracy (color-coded) with three parties (two benign and one malicious). \textsc{FedAvg} is located at the bottom left corner 
  ; QV is positioned around the centre 
  ; \textsc{FedQV} is situated along the right triangle side
  . Details of the experimental setup are provided in Appendix~\ref{ap:preliminary}.}
\label{fig:toy_model}
\end{wrapfigure}
\paragraph{Our contributions.}
In this paper, we propose \textsc{FedQV}, a novel FL aggregation scheme that draws inspiration from quadratic voting.
Our objective is to mitigate the ability of malicious peers who may have, or falsely claim to have, large datasets, to impose a disproportional damage on the global model -- a vulnerability inherent in the \textsc{FedAvg} that applies the \emph{1p1v} principle at the granularity of individual votes.
First, we demonstrate that the incorporation of QV into the FL setting restricts the ability of malicious peers to inflict high damages by taxing their credit voices more than linear. 
Figure~\ref{fig:toy_model} illustrates a toy use case with two benign and one malicious party engaged in a poisoning attack, with the dataset sizes set to $\left \{ 1,1,2 \right \}$, respectively. In contrast to \textsc{FedAvg}, which allocates aggregation weights as $\left \{1,1,2\right\}$, QV allocates weights as $\left\{ 1,1,\sqrt{2} \right \}$, effectively limiting the malicious party's influence.

To capture each party's preference for voting and enhance the detection of malicious updates, we require parties to submit the similarity of their local model with the previous round global model as their aggregation weight. Furthermore, in response to potential malicious attempts, we also introduce a truthfulness mechanism, \textsc{FedQV}, to our application of QV. This mechanism employs a masked voting rule and a limited budget to hide the vote calculation process from parties, preventing them from knowing the exact votes they have cast. These measures act as a deterrent against parties providing false information 
to evade penalties, which may exclude them from the current and following rounds.
Returning to our previous toy example, 
\textsc{FedQV} results in the allocation of weights $\left\{ 1,1,0 \right \}$, as illustrated in Figure~\ref{fig:toy_model}, that effectively excludes the malicious party from the aggregation, thereby increasing the accuracy of the resulting global model.

In election-related applications, QV allocates equal budgets to all voters, reflecting the democratic principle of equal rights. However, in our adaptation of QV for FL, it makes sense to allocate more votes to 
benign peers and limit the influence of malicious ones. We achieve this by employing unequal budgets, which are tied to a reputation score for each peer, as discussed in Section~\ref{subsec:adaptivebudget}. 
Furthermore, we design \textsc{FedQV} such that it can be easily combined with existing privacy-guaranteed mechanisms to thwart \emph{inference and reconstruction attacks}~\cite{melis2019exploiting,zhu2019deep}.

In terms of theoretical contributions, we present an extensive analysis in order to: 1) establish convergence guarantees, and 2) prove the truthfulness of our method.
We also conduct a thorough experimental evaluation for studying the accuracy, convergence, and resilience of our proposed mechanism against state-of-the-art Byzantine attacks on multiple benchmark datasets.

Our final contribution 
lies in extending the versatility of our core \textsc{FedQV} scheme, by facilitating its seamless integration with state-of-the-art Byzantine-robust FL defences.
This enables \textsc{FedQV} to serve as a complementary component, ultimately boosting the robustness of these existing defences, rather than being seen as a competitor. Notably, implementing these defences atop \textsc{FedQV} consistently yields superior results compared to employing them on top of \textsc{FedAvg}.


 


\paragraph{Our findings.} Using a combined theoretical and experimental evaluation, we show that:

\textbullet\ \textsc{FedQV} is a truthful mechanism and is theoretically and empirically compatible with \textsc{FedAvg} in terms of accuracy and convergence under attack and no-attack scenarios. 

\textbullet\ \textsc{FedQV} consistently outperforms \textsc{FedAvg} under various SOTA poisoning attacks, especially for local model poisoning attacks improving the robustness to such attacks by a factor of at least $4\times$.

\textbullet\  The combination of \textsc{FedQV} with a reputation model to assign unequal credit voice budgets to parties according to their respective reputations, improves robustness against poisoning attacks by at least 26\% compared to the baseline \textsc{FedQV} that uses equal budgets. 

\textbullet\ We show that integrating \textsc{FedQV} with established Byzantine-robust FL defences, including Multi-Krum~\cite{blanchard2017machine}, Trimmed-Mean~\cite{yin2018byzantine}, and Reputation~\cite{chu2022securing}, results in substantial enhancements in accuracy and reductions in the attack success rate under state-of-the-art attacks when compared to the original defence methods. 

\section{Related Work}
\subsection{Election Mechanisms in FL}
Election mechanisms are widely used in distributed systems for choosing a coordinator from a collection of processes~\cite{garcia1982elections,alford1985distributed}. 
Likewise, there exist works that explore the value of the election mechanism for the aggregation step of FL. Plurality voting is employed in \textit{FedVote}~\cite{yue2021federated} 
and \textit{FedVoting}~\cite{liu2021fedvoting} for treating the validation results as votes to decide the optimal model.
Also in~\cite{sohn2020election}, the authors propose two forms of election coding 
for discovering majority opinions for the aggregation step. \textit{DETOX}~\cite{rajput2019detox} proposes a hierarchical aggregation step based on majority votes upon groups of updates. Finally, \textit{DRACO}~\cite{chen2018draco} 
and \textit{ByzShield}~\cite{konstantinidis2021byzshield} also employ majority voting to fend off attacks against the aggregation step.
All the aforementioned election mechanisms suffer from the tyranny of the majority  problem in election systems~\cite{sartori1987theory}. In FL, this means that if attackers manage to control the majority of votes, then via poisoning their tyranny will manifest itself as a degradation of the accuracy of the FL model used by the minority.

To address these limitations,  QV is proposed as a solution that combines simplicity, practicality, and efficiency under relatively
broad conditions.
QV considers a quadratic vote pricing rule, inspired by economic theory, under which voters can purchase votes at ever-increasing prices within a predetermined voting budget. 
The advantages of QV over \emph{1p1v} have a rigorous theoretical basis, which of course applies also to the use of QV in FL. For any type of symmetric Bayes-Nash equilibrium, the price-taking assumption approximately holds for all voters, as a result, the expected inefficiency of QV is bounded by constant~\cite{lalley2016quadratic}.
This theoretical analysis~\cite {chandar2019quadratic,tideman2017efficient} combined with strong empirical validation, both at the laboratory~\cite{casella2019storable} and on the field~\cite{quarfoot2017quadratic}, suggest that QV is near-perfectly efficient and more robust than \emph{1p1v} which, as already explained, forms the basis of contemporary FL aggregation mechanisms.
The advantages of QV can also be observed from the viewpoint of collusion, which is generally deterred either by unilateral deviation incentives or by the reactions of non-participants~\cite{weyl2017robustness}. 


\subsection{Byzantine-robust FL Aggregation Against Privacy Attacks} There exist several Byzantine-robust FL aggregation methods for mitigating Byzantine attacks either by leveraging statistic-based outlier detection techniques~\cite{blanchard2017machine,yin2018byzantine,xie2019zeno,chu2022securing} or by utilising auxiliary labelled data collected by the aggregation server in order to verify the correctness of the received gradients~\cite{guo2021siren,cao2021fltrust}. Both approaches, though, require examining the properties of the updates of individual parties, which can jeopardise their privacy due to inference~\cite{melis2019exploiting} and reconstruction attacks~\cite{zhu2019deep,geiping2020inverting} mounted by an honest but curious aggregation server.
Contrary to those approaches, in \textsc{FedQV} the analysis of local updates and the calculation of corresponding weights is done locally at the peers using provably truthful mechanisms. This allows for the implementation of \textsc{FedQV} 
effectively using cryptographic techniques, such as \textit{the secure aggregation scheme}~\cite{bonawitz2016practical} and \textit{Fully Homomorphic Encryption}~\cite{aono2017privacy},
without being exposed to inference and reconstruction attacks at the aggregation server.  
It is worth noting that while there are alternative privacy-guaranteed mechanisms available in FL, such as differential privacy~\cite{dwork2006differential,du2019robust} and model inversion~\cite{zhao2022fedinv}, they do not provide the same level of security as cryptology-based methods~\cite{zhu2019deep}.
However, it is important to acknowledge that cryptographic methods are typically suitable for simple and specific computations like weighted averaging in \textsc{FedAvg} and \textsc{FedQV}. Hence, these methods are not applicable to more complex computations and data analyses required for the most Byzantine-robust FL aggregations. 
Although a few other FL aggregation approaches~\cite{ma2022privacy, so2020byzantine} can be adapted to incorporate cryptographic techniques, they still rely on majority voting as the aggregation scheme, which can be seamlessly integrated with \textsc{FedQV} to enhance its robustness against Byzantine attacks.

\section{Methodology}
\subsection{Federated Learning Setting}
Consider an FL system involving $N$ parties and a central server. During training round $t$, a subset of parties $\mathcal{S}^{t}$ is selected to participate in the training task. Party $i$ has the local dataset $\mathcal{D}_{i}$ with $\left | \mathcal{D}_{i} \right  |$ samples (voters), drawn from non-independent and non-identically (Non-IID) distribution $\mathcal{X}_{i} (\mu_{i},\,\sigma_{i}^{2})$. 
The goal of using FL is to learn a global model 
for the server. 
Given the loss function $\ell(\bm{w};\mathcal{D})$, 
the objective function of FL can be described as
    $\mathcal{L}(\bm{w}) = \mathbb{E}_{\mathcal{D}\sim \mathcal{X}}\left[\ell({\bm{w};\mathcal{D}}) \right]$.
Therefore, the task becomes:
    $ \bm{w}^{*} = \mathop{\arg\min}_{\bm{w} \in \mathds{R}^{d}} \mathcal{L}(\bm{w})$.
To find the optimal $\bm{w}^{*}$, Stochastic Gradient Descent (SGD) is employed to optimise the objective function. Let $T$ be the total number of every part’s SGD, $E$ be the local iterations between two communication rounds, and thus $\frac{T}{E}$ is the number of communication rounds. 

The FL model training process entails several rounds of communication between the parties and the server, including broadcasting, local training, and aggregation, as demonstrated in Algorithm~\ref{al:aggregtaion}. 
For aggregation rule, \textsc{FedAvg} uses the fraction of the local training sample size of each party over the total training samples as the weight of a party:
${\bm{w}}^{t+1} = \sum_{i\in\mathcal{S}^{t}}\left |\mathcal{D}_{i}  \right |\cdot \bm{w}_{i}^{t} / {\bigcup_{i\in \mathcal{S}^{t}}\left |\mathcal{D}_{i}  \right |}$.
Similar to \emph{1p1v}, each sample here represents a single voter, and since party $i$ possesses $\left |\mathcal{D}_{i}  \right |$ samples, it is able to cast $\left |\mathcal{D}_{i}  \right |$ votes for its local model during the aggregation. Hence, the global proposal is a combination of all parties' local proposals weighted by their votes.

\subsection{\textsc{FedQV}: Quadratic Voting in FL}
We use QV in FL to overcome the drawback of \emph{1p1v}, which improves the robustness of aggregation in comparison to \textsc{FedAvg} without compromising any efficiency. Our QV-based aggregation algorithm consists of two key components: \emph{similarity computation} and \emph{voting scheme}. 



\underline{\textbf{Similarity Computation:}}
In round $t$, based on the server instructions, party $i$ ($i\in\mathcal{S}^{t}$) trains its local model $ \bm{w}_{i}^{t}$, 
which can be regarded as its local proposal. 
Following the local training phase, party $i$ computes a similarity score $s^t_i$ utilising cosine similarity, quantifying the alignment between its locally trained model $\bm{w}{i}^{t}$ and the previous global model $\bm{w}^{t-1}$. 
Notably, the cosine similarity function can be adapted to different similarity metrics, such as L2 distance, to better suit specific tasks. In this context, a higher $s^t_i$ value indicates a stronger agreement with the previous global model(proposal).
Once selected parties finish training, they send their updates $\bm{w}_{i}^{t}$ to the server, with the message containing  $s^{t}_{i}$ and $\mathcal{D}_i$. 

Notice, that similarity calculation can also be launched on the server side, but it will (i) open the door to privacy attacks launched by the server, (ii) get the distorted similarity score due to some regularisation and privacy-preservation methods~\cite{mcmahan2017learning,mcmahan2018learning} employed in the party side. Since our main goal is to compare FedQV with FedAvg, in which the weights are also calculated on the party side, in FedQV we set the calculation on the party side. On the server side we can add the defence layers such as other byzantine-robust aggregations as we show in Section XX, the attacker is harder to attack successfully than FedAvg.

\underline{\textbf{Voting Scheme (Server Side):}} 
Upon receiving the updates and messages from selected parties, the server proceeds with the following steps:

(i) The server normalises the similarity scores $s_{i}^{t}$ using Min-Max Scaling to obtain $\bar{s}_{i}^{t} \in [0,1]$; 

(ii) The server penalises parties with abnormal similarity scores ($\bar{s}_{i}^{t}\leq \theta$ or $\bar{s}_{i}^{t}\geq 1-\theta$), where $\theta$ is the similarity threshold. This addresses excessively large or small similarity scores, which are considered suspicious. Penalties are applied by adjusting their budget $B_i$ with the formula: $B_i = \max \left ( 0, B_i + \ln{\bar{s}_{i}^{t}}-1 \right )$;

(iii) The server calculates the voice credit $c_{i}^{t}$ for party $i$ utilising the masked voting rule $\mathcal{H}$:
\begin{equation}
\label{eq:credit}
c_{i}^{t} = \mathcal{H}(\bar{s}_{i}^{t}) = \left ( -\ln{\bar{s}_{i}^{t}}+1 \right )\mathds{1}_{\theta < \bar{s}_{i}^{t}< 1-\theta} 
\end{equation}
Here, the voice credit signifies the price party $i$ is required to pay in round $t$ for its local proposal.
Parties with higher similarity scores, indicating stronger agreement with the global proposal, receive fewer credit votes from the server. 
(iv) The server checks the budget $B_i$ for each party $i$ and employs QV to compute their final votes $v_{i}^{t}$:
\begin{equation}
\label{eq:vote}
   v_{i}^{t} = \sqrt{\min \left (\left | \mathcal{D}_{i} \right |c_{i}^{t}, \max \left ( 0, B_{i}\right )\right )}
\end{equation}
Subsequently, the server updates the budget as follows:
$B_i = \max (0, B_i - \left(v_{i}^{t}\right)^2)$.

Thus, the server determines the weight ($v_{i}^{t}$) of party $i$ for aggregation and generates the updated global model $\bm{w}^{t-1}$. 
Algorithm~\ref{al:aggregtaion} summarises all these steps of \textsc{FedQV}.
 
In cases where parties attempt to manipulate the similarity scores, their power is constrained by: 

\emph{(i) No knowledge of the voting process.} Only the server possesses knowledge of each party's remaining budget and the number of actual votes cast in the current round.
This feature ensures that parties remain unaware of the inner workings of the credit voice allocation process for aggregation. Consequently, even if parties possess a comprehensive understanding of how \textsc{FedQV} functions on the server side, they remain incapable of strategizing or predicting their credit voice allocation. 

\emph{(ii) Punitive Measures}: \textsc{FedQV}, with its masked voting rule and limited budget, has provisions to penalise and remove malicious participants, acting as a strong deterrent against manipulation attempts; 

\emph{(iii) Limited Influence}: Even if a manipulated similarity score is accepted by the server, the influence the malicious participant can exert is inherently constrained due to the nature of QV, minimising the potential damage. 


\begin{algorithm}[!t]
   \SetKwInOut{Input}{Input}
   \SetKwInOut{Output}{Output}
   \SetKwInOut{Server}{Server}
   \SetKwInOut{Party}{Party}
   \SetKwFunction{Norm}{Norm}
   \SetKwFunction{$S_{C}$}{$S_{C}$}
   \SetKwFunction{H}{$\mathcal{H}$} 
   \SetAlgoLined
    \Input{
    $\bm{w}^{0}$ $\leftarrow$ random initialisation; $B$, $\theta$ $\leftarrow$ \textsc{FedQV} parameters}
    \BlankLine
    \Server{}
     \For{Iteration $t\leftarrow 1$ \KwTo $\frac{T}{E}$}
    {
     Broadcast $\bm{w}^{t-1}$ to randomly selected set of parties $\mathcal{S}^{t}$ ($\left | \mathcal{S}^{t} \right | = \mathcal{C} \geq 1$);\\
     Receive \emph{the local updates} $(\bm{w}^{t},s^{t},\left | \mathcal{D} \right |)$ from selected parties ($i\in\mathcal{S}^{t}$) and compute the normalised $\bar{s}_{i}^{t}$ ;\\
    \ForPar{$i\leftarrow 1$ \KwTo $N$}{
     \uIf{ $\bar{s}_{i}^{t}\leq \theta$ or $\bar{s}_{i}^{t}\geq 1-\theta$}{ Update $B_i$ $\leftarrow \max \left ( 0, B_i + \ln{\bar{s}_{i}^{t}}-1 \right )$}
     \emph{Credit voice}  $c_{i}^{t}\leftarrow$ Equation~\ref{eq:credit} and \emph{Vote} $v_{i}^{t} \leftarrow$ Equation~\ref{eq:vote};\\
     \emph{Budget}  
     $B_i$ $\leftarrow \max \left(0, B_i - \left(v_{i}^{t}\right)^2 \right ) $
     }
     \Return $\bm{w}_{n}^{t}  \leftarrow \sum_{i=1}^{N}\frac{v_{i}^{t}}{\sum_{i=1}^{M} v_{i}^{t}}\bm{w}_{i,n}^{t}$
   } 
    \BlankLine
    \Party{}
    \setcounter{AlgoLine}{0}
    \ForPar{Party $i \in \mathcal{S}^{t}$}{
    Receive \emph{the global update} $\bm{w}^{t-1}$ and conduct local training $\bm{w}_{i}^{t} \leftarrow  \bm{w}_{i}^{t-1} - r_{t-1}\frac{\partial \ell_{i}({\bm{w}_{i}^{t-1};\mathcal{D}_{i}})}{\partial \bm{w}}$;\\
     Calculate \emph{the similarity score} $s_{i}^{t} \leftarrow$
      {$\frac{\left \langle \bm{w}_{i}^{t},  \bm{w}^{t-1} \right \rangle}{\left \|  \bm{w}_{i}^{t}  \right \|\cdot\left \| \bm{w}^{t-1} \right \|}$} and send back $\left(\bm{w}_{i}^{t},s_{i}^{t},\left | \mathcal{D}_{i} \right |\right)$\\
    \BlankLine
     }
\caption{\textsc{FedQV}}
\label{al:aggregtaion}
\end{algorithm}

\paragraph{{Benefits of \textsc{FedQV}}}
\textbf{1. Truthful Mechanism.} 
\textsc{FedQV} is a \emph{truthful} mechanism~\cite{blumrosen2007algorithmic} as we prove in Theorem~\ref{th:theorem2}. This means that this mechanism compels the parties, even malicious ones, to tell the truth about their votes (weights) for aggregation, rather than any possible lie. This truthfulness is reinforced by the aforementioned several defence layers.
\textbf{2. Ease of Integration and Compatibility.} 
\textsc{FedQV} is highly adaptable and can be seamlessly integrated into Byzantine-robust FL defence schemes with minimal adjustments, specifically by modifying the aggregation weight calculation while leaving other algorithm components unchanged. This integration is demonstrated in Section~\ref{subsec:krum}. 
Furthermore, similar to \textsc{FedAvg}, \textsc{FedQV}  boasts efficient communication and simplicity, rendering it compatible with various mechanisms employed in FL. It can effortlessly incorporate the regularisation, sparsification, and privacy modules, encompassing techniques such as clipping~\cite{mcmahan2017learning}, gradient compression~\cite{sattler2019sparse}, differential privacy~\cite{dwork2006differential}, and secure aggregation~\cite{bonawitz2016practical}. 

\vspace{-2pt}
\subsection{\textsc{FedQV} with Adaptive Budgets}
In democratic elections, all individuals are typically granted equal voting rights, entailing an equal voting budget. In FL, however, it often makes sense to give malicious parties fewer votes than honest ones. Thus to improve the robustness of standard \textsc{FedQV}, we combine it with the reputation model in~\cite{chu2022securing} to assign an unequal budget based on the reputation score of parties in each round $t$. 
Specifically, if a party's reputation score $R^{t}$ surpasses a predefined threshold $\lambda$, we increase their budget, and vice versa. 
We present a summary of this combination in Algorithm~\ref{al:adaptive_bugdet}, with a detailed explanation provided in Appendix~\ref{ap:algo_budget}, expanding on the well-established components from the original paper.
We provide empirical evidence in section~\ref{sec:experiments} showcasing the substantial performance improvements achieved by the enhanced version of \textsc{FedQV} featuring an adaptive budget.

\vspace{10pt}
\begin{algorithm}[!htbp]
   \SetKwInOut{Input}{Input}
   \SetKwInOut{Output}{Output}
   \SetKwInOut{Server}{Server}
   \SetKwFunction{Rep}{Rep}
   \SetKwFunction{IRLS}{IRLS}
   \SetKwFunction{ReLU}{ReLU}
    \SetAlgoLined
     \Input{$\bm{w}_{i}^{t} , c_{i}^{t}, B_{i}^{t}$ $\leftarrow$ \textsc{FedQV}; $\kappa$,$a$,$W$,$M$,$\lambda$,$\delta$$\leftarrow$ \emph{Reputation model} parameters}
    \BlankLine
     \For {$i\in \mathcal{S}^{t}$}{
     \For {$ j \leftarrow 1$ \KwTo $M$}{
     \emph{Subjective Observations} $(P_{i}^{t},N_{i}^{t}) := $ \IRLS ($w_{i,j}^{t},\delta$)\;}
     \emph{Reputation Score} $R_i^t := $ \Rep $(P_{i}^{t},N_{i}^{t},\kappa,a,W)$\\
     \emph{Budget} $B_i^t \leftarrow R_i^t\mathds{1}_{\lambda \leq R_i^t} + B_i^t$
     , \emph{Credit voice} $c_i^{t} \leftarrow (R_i^t + c_i^{t})\mathds{1}_{\lambda \leq R_i^t}$
      \BlankLine }
\caption{FedQV with Adaptive Budget}
\label{al:adaptive_bugdet}
\end{algorithm}

\section{Theoretical Analysis}
In this section, we show that the convergence for \textsc{FedQV} is guaranteed in bounded time and that \textsc{FedQV} is a truthful mechanism. Our first major result is Theorem~\ref{th:theorem_malicious} that states \textsc{FedQV} converges to the global optimal solution at a rate of $\mathcal{O}(\frac{1}{T})$, where $T$ is the total number of SGD, for strongly convex and smooth functions with non-iid data. Regarding the performance of our algorithm in terms of metric average accuracy and convergence as will be illustrated in the following section, we show that it is consistent with our theoretical analysis. Our second major result is Theorem~\ref{th:theorem2} which states that \textsc{FedQV} is a truthful mechanism. Fully detailed proofs are provided in the Appendix~\ref{sec:th_proof}.

\subsection{Convergence}

Suppose the percentage of attackers in the whole parties is $m$, we denote
$$\mathcal{M}_{i}(\bm{w}_{i}^{t}) = \begin{cases}
* & \text{ if } i \in {\mbox{malicious parties}} \\ 
\nabla \ell(\bm{w}_{i}^{t};\mathcal{D}_{i}^{t}) & \text{ if } i \in {\mbox{honest parties}}
\end{cases}$$
Where $*$ stands for an arbitrary value from the malicious parties. Under four mild and standard assumptions for such types of analysis in accordance with recent works~\cite{yin2018byzantine, xie2019zeno, yu2019parallel,cao2021fltrust,chu2022securing,cao2022fedrecover}, along with the support of Lemmas outlined in the Appendix~\ref{ap:lemmas}, we have 
\begin{theorem}
\label{th:theorem_malicious}
Under Assumptions \ref{as:assumption1}, \ref{as:assumption2}, \ref{as:assumption3} and~\ref{as:assumption4}, Choose $\alpha = \frac{L+\mu}{\mu L}$ and $\beta =  2\frac{(L+1)(L+\mu)}{\mu L}$, then \textsc{FedQV} satisfies
\begin{align}
     \EX \mathcal{L}(\bm{w}^T) - \mathcal{L}(\bm{w}^*) \leq \frac{L + 2Lr_{T-1}\varpi}{2\varphi + T}\left( \varphi\EX\left \|\bm{w}^{0}
      - \bm{w}^{*} \right \|_{2}^{2} + \frac{\alpha^{2}}{2}\Delta \right) 
     + \frac{L\varpi^2}{2}
\end{align}
Where
\begin{gather*}
    \Delta = \left(E-1\right)^{2} \mathcal{G}_{\bm{w}}^{2}
    + \left(1-2\theta\right)\mathcal{C}\mathcal{V}_{\bm{w}}\sqrt{B},\; 
    \varphi = \alpha \left ( L+1  \right),\; \varpi = mN\mathcal{G}_{\bm{w}}r_{T-1}\sqrt{4 + 6\theta -\theta^2}
\end{gather*}
\end{theorem}
\begin{remark}
\label{re:remark1}
According to Theorem~\ref{th:theorem_malicious} and Theorem~\ref{th:theorem1}, \textsc{FedQV} obtains a convergence rate of $\mathcal{O}(\frac{1}{T})$ irrespective of the presence or absence of adversarial participants, which is comparable to the convergence rate of \textsc{FedAvg}~\cite{li2019convergence}.
\end{remark}

\begin{remark}
\label{re:remark2}
The error rate exhibits dependence on the budget $B$, the similarity threshold $\theta$, and the percentage of malicious parties $m$. It is noteworthy that a larger budget allocation, a reduction in the similarity threshold, or an augmentation in the proportion of malicious parties induce more pronounced disparities in model updates, consequently resulting in an elevated error rate. The impact of these hyperparameters is shown in Figure~\ref{fig:hyper} in the Appendix~\ref{ap:hyper}.
\end{remark}

\subsection{Truthfulness}
The \textsc{FedQV} mechanism belongs to a single-parameter domain since the single real parameter votes $v_{i}$ directly determines whether party $i$ will be able to join
the aggregation. 
In addition, it is normalised according to the definition in the game theory~\cite{blumrosen2007algorithmic} that for every $v_{i}$, $v_{-i}$ such that $f(v_{i},v_{-i})\notin W_{i}$, $p_{i}(v_{i},v_{-i}) = 0$. Here, $v_{-i}$ denotes the votes cast by all other parties except for $i$, 
$W_{i}$ represents the subset of participants in aggregation, $f$ is the outcome of the voting scheme, and $p_i$ is the payment function that $p_i(v_{i},v_{-i}) = v_{i}^{2}$ in \textsc{FedQV}. The following is the definition of truthfulness and lemmas that we use in the proof of the Theorem~\ref{th:theorem2} in accordance with monotone and critical value in the game theory~\cite{blumrosen2007algorithmic}. 
\begin{definition}
A mechanism $(f, p_{1},...,p_{n})$ is called truthfulness if for every party $i$, we denote $a = f (v_{i},v_{-i})$ and $a' = f (v^{'}_{i},v_{-i})$ as the outcome of the voting, then $v_i(a) - p_{i}(v_{i},v_{-i}) \geq v_i(a^{'}) - p_{i}(v^{'}_{i}, v_{-i})$, where $v_i(a)$ denotes the gain of party $i$ if the outcome of the voting is $a$. 
\end{definition}
$v_i(a) - p_{i}(v_{i},v_{-i})$ is the utility of party $i$, which means the gain from voting ($v_i(a)$) minus its cost ($p_i(v_i,v_-i)$). Intuitively this means that party $i$ with $v_{i}$ would prefer “telling the truth” $v_{i}$ to the server rather than any possible “lie” $v^{'}_{i}$ since this gives him higher (in the weak sense) utility.

\noindent Based on Lemma~\ref{le:monotone} and~\ref{le:critical value} in Appendix~\ref{le:lemma_trustful}, we have:
\begin{theorem}
\label{th:theorem2}
\textsc{FedQV} is incentive compatible (truthful).
\end{theorem}


\begin{remark}
\label{re:remark3}
Regarding the concept of truthfulness, it theoretically ensures that being honest is the dominant strategy since providing manipulated similarity scores may lead to penalties and removal from the system due to the masked voting rule $\mathcal{H}$ and limited budget $B$. This is an integral part of the nature of QV embedded within our \textsc{FedQV} framework.
\end{remark}

\section{Experiments}
\label{sec:experiments}
\subsection{Experimental setting}
\paragraph{Datasets and global models}
We implement the typical FL setting where each party owns its local data and transmits/receives information to/from the central server. To demonstrate the generality of our method, we train different global models on different datasets. We use four popular benchmark datasets: MNIST~\cite{lecun1998mnist}, Fashion-MNIST~\cite{xiao2017fashion}, FEMNIST~\cite{caldas2019leaf} and CIFAR10~\cite{krizhevsky2009learning}. We consider a multi-layer CNN same as in ~\cite{mcmahan2017learning} 
for MNIST, Fashion-MNIST and FEMNIST, and the ResNet18~\cite{he2016deep} 
for CIFAR10.
\vspace{-3pt}
\paragraph{Non-IID setting} In order to fulfil the setting of a heterogeneous and unbalanced dataset for FL, we sample 
from a Dirichlet distribution with the concentration parameter $\iota = 0.9$ as the Non-IID degree as in~\cite{bagdasaryan2020backdoor, hsu2019measuring}, 
with the intention of generating non-IID and unbalanced data partitions.
Moreover, we have examined the performance across varying levels of non-IID data, spanning from 0.1 to 0.9, as depicted in Appendix~\ref{ap:non-iid}.
\vspace{-3pt}
\paragraph{Parameter Settings}
The server selects 10 ($\mathcal{C}$) out of 100 ($N$) parties to participate in each communication round and 
train the global models for 100 communication rounds($\frac{T}{E}$). We set the model hyper-parameters budget $B$ and the similarity threshold $\theta$ to 30 and 0.2 respectively based on the hyper-parameter searching.
All additional settings are provided in the Appendix~\ref{sec:setting}. 

\subsection{Evaluated Poisoning Attacks}
Our paper addresses two distinct attack schemes:

\textbullet\ \textbf{Data poisoning}: Attackers submit the true similarity score based on their poisoned updates, including \textbf{Labelflip Attack}~\cite{fang2020local}, \textbf{Gaussian Attack}~\cite{zhao2022fedinv},  \textbf{Backdoor }~\cite{gu2019badnets}, \textbf{Scaling Attack}~\cite{bagdasaryan2020backdoor},  \textbf{Neurotoxin }~\cite{zhang2022neurotoxin}.

\textbullet\ \textbf{Model poisoning}: Attackers submit the true similarity score based on their clean updates and poison their model, including: \textbf{Krum Attack}~\cite{fang2020local}, \textbf{Trim Attack}~\cite{fang2020local}, and    Aggregation-agnostic attacks: \textbf{Min-Max} and \textbf{Min-Sum}~\cite{shejwalkar2021manipulating}

Moreover, we introduce an adaptive attack, \textbf{QV-Adaptive}, tailored for \textsc{FedQV}, leveraging the AGR-agnostic optimisations~\cite{shejwalkar2021manipulating} within the LMP framework~\cite{fang2020local} to manipulate both the similarity score and the local model.

The details of these attacks are in Appendix~\ref{ap:attacks}. It is noteworthy that  Labelflip, Gaussian, Krum, Trim, Min-Max, Min-Sum and QV Adaptive attacks are untargeted attacks, whereas, Backdoor, Scaling and Neurotoxin attacks are targeted attacks.
We confine our analysis to the worst-case scenario in which the attackers submit the poisoned updates in every round of the training process for all attack strategies with the exception of the Scaling attack. 

\subsection{Performance Metrics}
\vspace{-6pt}
We use the average test accuracy (ACC) of the global model to evaluate the result of the aggregation defence for poisoning attacks. 
In addition, there are targeted attacks that aim to attack a specific label while keeping the accuracy of classification on other labels unaltered. Therefore, besides ACC, we choose the attack success rate (ASR) to measure how many of the samples that are attacked, are classified as the target label chosen by malicious parties.
\begin{figure}
    \centering
    \begin{subfigure}[t]{0.48\textwidth}
    \centering
    \includegraphics[width=\columnwidth]{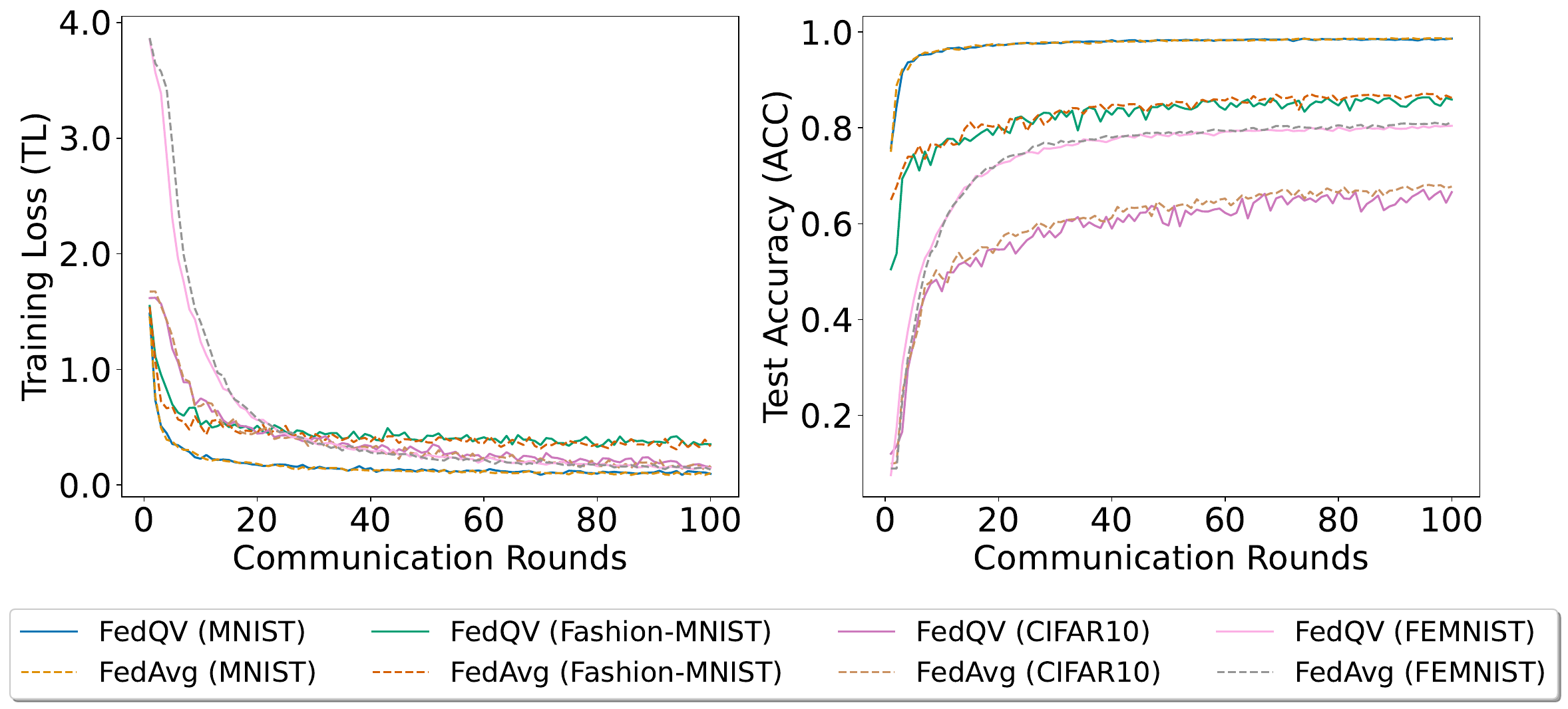}
    \caption{Training Loss and ACC for 100 epochs of \textsc{FedQV} and \textsc{FedAvg} in four benchmark datasets under no attack scenario.}
    \label{fig:convergence}
\end{subfigure}
\hfill
    \begin{subfigure}[t]{0.48\textwidth}
    \centering
    \includegraphics[width=\columnwidth]{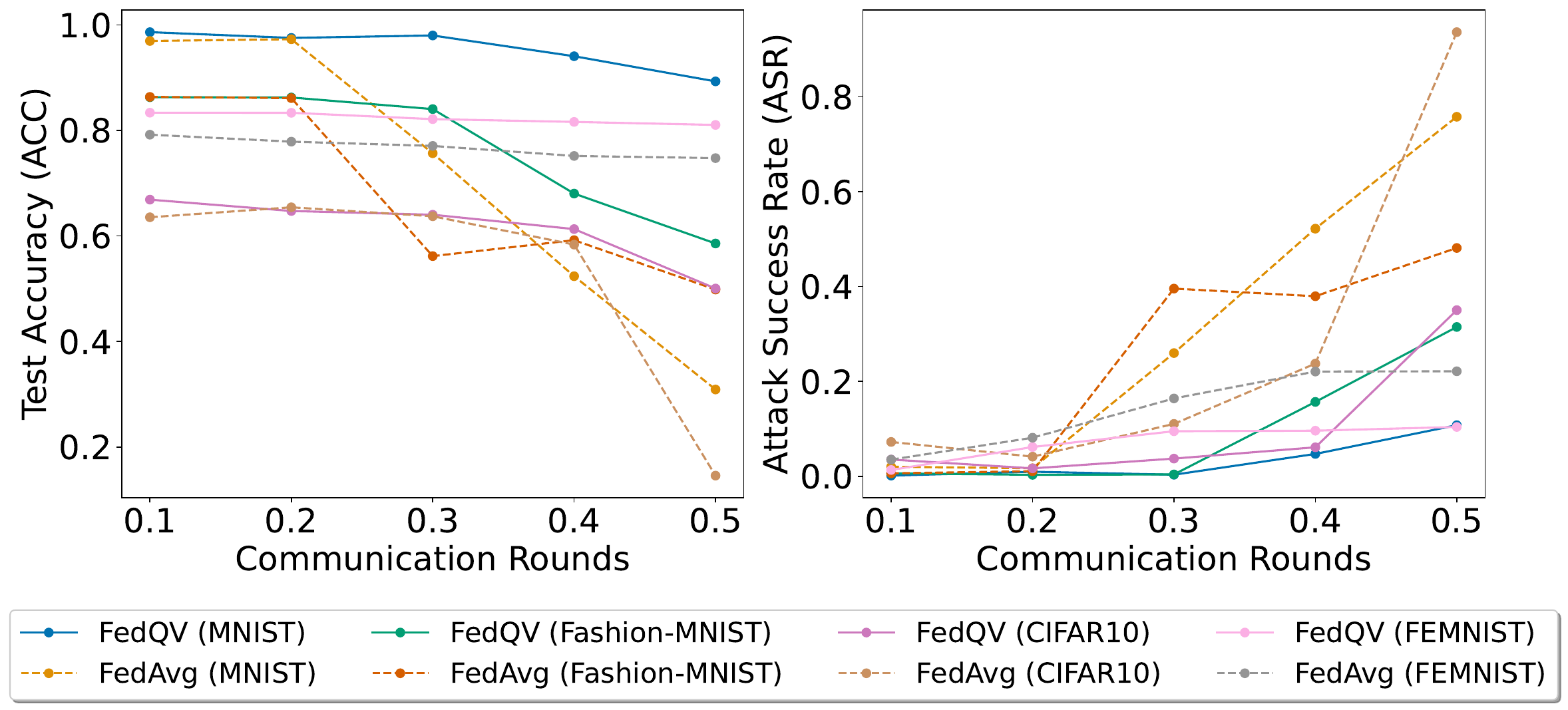}
    \caption{ACC and ASR for 100 epochs of \textsc{FedAvg}, \textsc{FedQV} in four benchmark datasets under Backdoor attack with varying $m$ from 10\% to 50\%.}
\label{fig:fedqv_backdoor}
    \end{subfigure}
\hfill
    \begin{subfigure}{0.46\textwidth}
    \centering
    \includegraphics[width=\columnwidth]{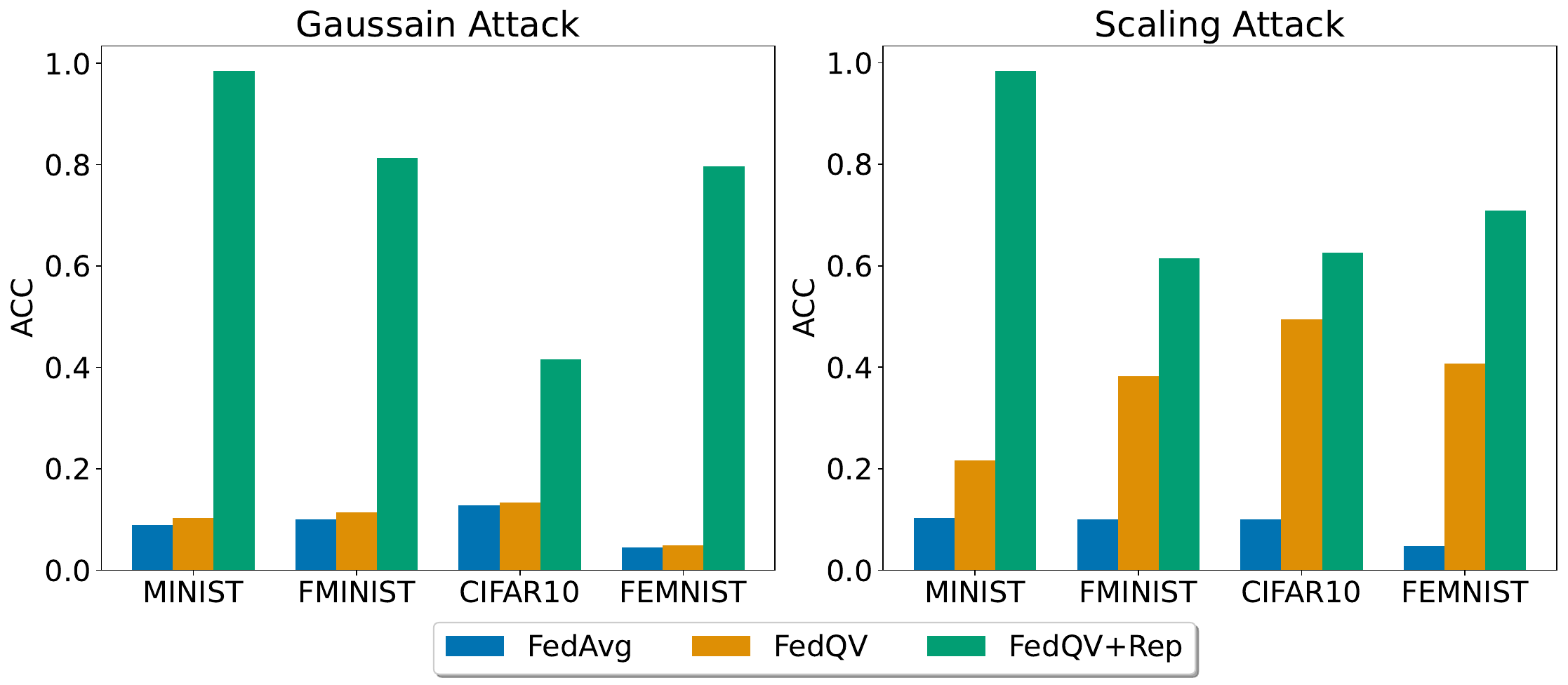}
    \caption{ACC for 100 epochs of \textsc{FedAvg}, \textsc{FedQV}, \textsc{FedQV} + \textsc{Rep}(\textsc{FedQV} with reputation model) in four benchmark datasets under 2 attack scenarios with 50\% malicious parties.}
    \label{fig:fedqv_rep}
    \end{subfigure}
     \hfill
     \begin{subfigure}{0.46\textwidth}
    \centering
    \includegraphics[width= \columnwidth]{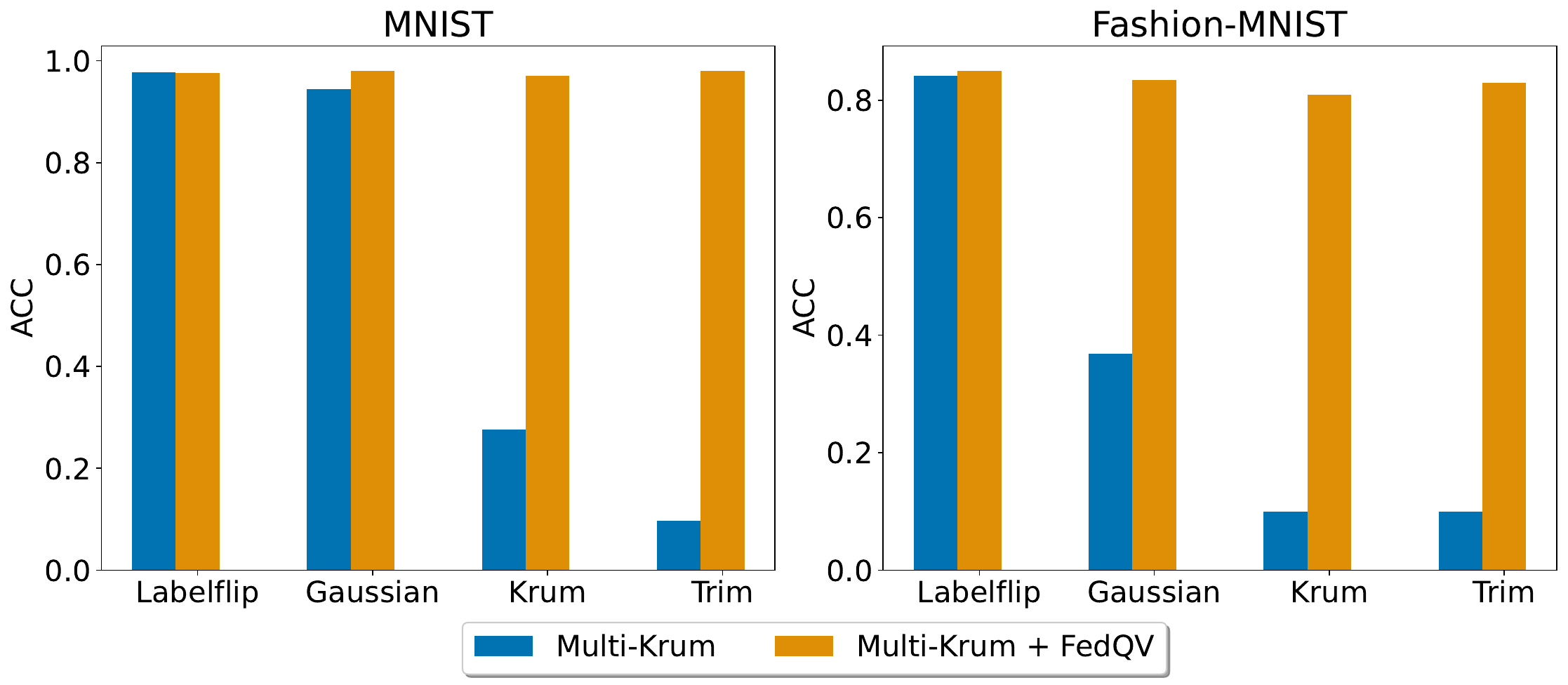}
    \caption{Average test accuracy for 100 epochs of Krum and Multi-Krum + \textsc{FedQV} on two benchmark datasets under 4 untargeted attack scenarios with 30\% malicious parties.}
    \label{fig:Krum}
\end{subfigure}
\vspace{-10pt}
\end{figure}

\subsection{Convergence}
\vspace{-6pt}
We evaluate the convergence of \textsc{FedAvg} and \textsc{FedQV} in the aforementioned four datasets without attack. 
We plot the training loss and ACC of the global models trained via \textsc{FedQV} and \textsc{FedAvg} in Figure~\ref{fig:convergence}.
We observe that, in the absence of Byzantine attacks, the global model trained using \textsc{FedQV} converges as fast as that under \textsc{FedAvg} for all four datasets, aligning with Theorem~\ref{th:theorem_malicious}.
\subsection{Defence against Poisoning Attacks}
\vspace{-6pt}
We present ACC and ASR results of global models trained using both \textsc{FedAvg} and \textsc{FedQV} under the 10 aforementioned attacks, with 30\% malicious parties for all four datasets, in Table~\ref{tab:attack}. In data poisoning attacks, the results consistently demonstrate that \textsc{FedQV} outperforms \textsc{FedAvg}, achieving the highest ACC with the smallest standard error. When considering targeted attacks, \textsc{FedQV} again stands out, displaying the highest ACC along with the lowest ASR when compared to \textsc{FedAvg}.
In the context model poisoning attacks, \textsc{FedQV} consistently outperforms \textsc{FedAvg}, except for the QV-Adaptive attack, which is tailored for \textsc{FedQV}. Especially for local model poisoning attacks: Trim and Krum attacks, \textsc{FedQV} outperforms \textsc{FedAvg} by at least \emph{4 times } in terms of accuracy.


Then we vary the percentage of attackers from 10\% to 50\% in Figure~\ref{fig:fedqv_backdoor} under the backdoor attack. Remarkably, \textsc{FedQV} outperforms the baseline regarding ACC and ASR across all scenarios, even when half the parties are malicious. 
To investigate  the behaviour of \textsc{FedQV} in scenarios with finer gradations, we also evaluate it with small, realistic percentages of attackers, same as in ~\cite{shejwalkar2022back}, in Table~\ref{tab:defenses} and Appendix~\ref{tab:target_small}. However, we notice that none of these methods
yields satisfactory accuracy results for Gaussian and Scaling attacks. To address this, we present the enhanced version of \textsc{FedQV} with an adaptive budget assigned according to a reputation model.

\begin{table}[t]
\centering
\resizebox{1\columnwidth}{!}{%
\begin{tabular}{@{}rrrcrrcrrcrrc@{}}
\toprule
& \multicolumn{2}{c}{MNIST} & \phantom{abc}& \multicolumn{2}{c}{Fashion-MNIST} &
\phantom{abc} & \multicolumn{2}{c}{CIFAR10} &
\phantom{abc} & \multicolumn{2}{c}{FEMNIST}\\
\cmidrule{2-3} \cmidrule{5-6} \cmidrule{8-9}  \cmidrule{11-12}& 
\textsc{FedAvg} & \textsc{FedQV} && \textsc{FedAvg} & \textsc{FedQV} && \textsc{FedAvg} & \textsc{FedQV} && \textsc{FedAvg} & \textsc{FedQV} 
\\ \midrule 
Data Poison\\
Labelflip &
  98.81$\pm$0.03 &
  98.54$\pm$0.05 &
   &
  \textbf{86.70$\pm$0.02} &
  85.22$\pm$0.05 &
   &
  66.88$\pm$0.48 &
  67.36$\pm$0.22 & &
  74.92$\pm$2.55& 
  \textbf{78.42$\pm$0.65}
  \\
Gaussian &
  9.68$\pm$0.41 &
  10.49$\pm$0.46 &
   &
  10.00$\pm$0.00 &
  \textbf{27.38$\pm$17.38} &
   &
  15.29$\pm$0.57 &
  \textbf{19.76$\pm$3.66} & &
  4.64$\pm$0.13&
  4.83$\pm$0.25\\
Backdoor\\
ACC(\%) &
  37.38$\pm$19.82 &
  \textbf{98.30$\pm$0.15} &&
  74.27$\pm$9.12 &
  \textbf{78.40$\pm$3.95} &&
  59.85$\pm$2.18 & 60.65$\pm$1.72&&
  49.78$\pm$22.38 &
  \textbf{75.20$\pm$3.96} 
  \\
ASR(\%) &
  68.49$\pm$22.00 &
  \textbf{0.19$\pm$0.07} &&
  14.58$\pm$12.53 &
  \textbf{7.05$\pm$6.35} &&
  18.20$\pm$5.27 &
  \textbf{3.21$\pm$1.30} &&
  30.88$\pm$7.52&
  \textbf{28.26$\pm$9.57}\\
Scaling\\
ACC(\%) &
  10.33$\pm$0.05 &
  11.16$\pm$0.88 &
   &
  10.22$\pm$0.09 &
  11.27$\pm$0.99 &
   &
  10.00$\pm$0.00 &
  \textbf{28.55$\pm$18.55}&
  &
  26.30$\pm$21.55&
  \textbf{64.80$\pm$1.38}\\
ASR(\%) &
  99.94$\pm$0.06 &
  98.96$\pm$1.04 &
   &
  99.74$\pm$0.10 &
  98.21$\pm$1.45 &
   &
  100.00$\pm$0.00 &
  67.66$\pm$32.34 &
  &
   0.47$\pm$0.08&
    0.56$\pm$0.06
  \\

Neurotoxin\\
ACC(\%) & 81.17$\pm$15.39
   & \textbf{95.73$\pm$1.45}
   && 70.00$\pm$7.85
   & \textbf{79.58$\pm$1.60}
   && 22.40$\pm$7.16
   & \textbf{45.40$\pm$3.22}
  && 47.29$\pm$18.07
  & \textbf{79.99$\pm$0.70}
  \\
ASR(\%) & 23.19$\pm$2.25
   & \textbf{18.11$\pm$1.67}
   && 20.65$\pm$2.21
   & 18.12$\pm$4.16
   && 51.63$\pm$1.03
   & 57.42$\pm$1.91
  && 40.42$\pm$4.35
   & \textbf{9.00$\pm$1.29}
   \\

  \midrule

Model Poison\\
Krum &
  10.57$\pm$0.39 &
  \textbf{97.96$\pm$0.14} &
   &
  10.00$\pm$0.00 &
  \textbf{79.43$\pm$0.86} &
   &
  10.00$\pm$0.00 &
  \textbf{53.27$\pm$1.12} & &
  5.20$\pm$0.22 
  &  \textbf{51.86$\pm$3.06}\\
Trim &
  10.04$\pm$0.16 &
  \textbf{98.36$\pm$0.11} &
   &
  10.00$\pm$0.00 &
  \textbf{84.45$\pm$0.70} &
   &
  10.00$\pm$0.00 &
  \textbf{57.33$\pm$2.34} & &
   5.09$\pm$0.33
   &  \textbf{52.19$\pm$4.52} \\

Min-Max & 35.00$\pm$25.38
   & \textbf{85.32$\pm$6.45}
  & 
   & 10.00$\pm$0.00
   &  \textbf{67.25$\pm$7.44} 
   &
   & 10.00$\pm$0.00
   & \textbf{19.07$\pm$6.97} 
   && 56.37$\pm$13.67
  &\textbf{72.58$\pm$2.11}
  \\

Min-Sum & 96.69$\pm$0.94
   & 95.97$\pm$0.59
  &
   & 10.88$\pm$0.87
   &  \textbf{83.93$\pm$0.81} 
   & 
   & 17.40$\pm$4.27
   & \textbf{43.94$\pm$3.56} 
   &&52.56$\pm$23.91
  &\textbf{72.36$\pm$1.61}
   \\

QV-Adaptive & \textbf{71.43}$\pm$22.67
   & 56.94$\pm$23.95
  &
   & 35.92$\pm$4.60
  &  \textbf{62.13$\pm$11.25} 
  & 
  & 10.00$\pm$0.00
  & 11.14$\pm$1.14
  & & 22.08$\pm$18.72
   & \textbf{43.78$\pm$20.72}
  \\

\bottomrule
\end{tabular}}
\vspace{2pt}
\caption{Comparison of \textsc{FedQV} and \textsc{FedAvg} on four benchmark Datasets under 10 attack scenarios with 30\% malicious parties. The Best Results are highlighted in bold.}
\label{tab:attack}
\vspace{-2pt}
\end{table}


\begin{table}[]
\centering
\resizebox{1\columnwidth}{!}{%
\begin{tabular}{@{}rcccccccc@{}}
\toprule
\multicolumn{1}{l}{} &
  Multi-Krum &
  Multi-Krum + FedQV &
   &
  Trimmed-Mean &
  Trimmed-Mean + FedQV &
   &
  Rep &
  Rep + FedQV \\ \midrule
Neurotoxin                       \\
1\%  & 78.99$\pm$1.03/1.05$\pm$0.01  & 80.61$\pm$0.66/0.86$\pm$0.17 &  & 85.23$\pm$1.77/0.59$\pm$0.13& 84.75$\pm$0.84/0.45$\pm$0.06  &  & 80.99$\pm$1.15/0.84$\pm$0.32 & \textbf{85.82$\pm$0.55/0.38$\pm$0.06}  \\
5\%  & 76.21$\pm$0.73/3.29$\pm$0.92  & 80.32$\pm$1.07/\textbf{1.34$\pm$0.34} &  & 85.15$\pm$0.38/0.87$\pm$0.17 &  85.09$\pm$0.97/0.73$\pm$0.05&  & 80.48$\pm$1.17/1.62$\pm$0.11 & \textbf{84.12$\pm$0.38/1.35$\pm$0.40}  \\
10\% & 72.79$\pm$1.02/21.73$\pm$7.07 & \textbf{77.41$\pm$1.36/16.30$\pm$3.01} &  & 85.13$\pm$0.70/2.32$\pm$0.27 &  84.68$\pm$0.73/1.45$\pm$0.21&  & 80.86$\pm$0.86/1.30$\pm$0.10 &  \textbf{83.78$\pm$0.09/0.66$\pm$0.04}  \\
Min-Max                               \\
10\% & 71.62$\pm$4.48 & \textbf{79.48$\pm$0.82} &  & 75.41 $\pm$0.77 & 78.46$\pm$0.67&  & 72.66$\pm$1.34 & \textbf{76.98$\pm$1.48} \\
30\% & 52.29$\pm$0.34 & \textbf{58.42$\pm$4.92} &  & 59.62 $\pm$1.20 &60.24$\pm$6.81 &  & 54.94$\pm$0.82 & \textbf{58.02$\pm$0.71} \\
50\% & 10.28$\pm$0.28 & \textbf{22.95$\pm$9.94} &  & 9.47$\pm$0.42 & 10.64$\pm$0.63 &  & 11.23$\pm$1.00 & 13.64$\pm$2.58 \\

QV-Adaptive                           \\
10\% & 52.98$\pm$1.78 & \textbf{73.35$\pm$3.44} &  & 83.17$\pm$1.85 & \textbf{85.55$\pm$0.33} &  & 12.60$\pm$2.36 & \textbf{41.55$\pm$19.78}  \\
30\% & 34.93$\pm$14.60 & \textbf{55.07$\pm$11.40} &  & 29.14$\pm$19.14 & \textbf{42.17$\pm$25.95} &  & 12.44$\pm$2.44 & \textbf{38.44$\pm$14.32} \\
50\% &  10.20$\pm$0.20 & 12.24$\pm$1.12  &  & 10.00$\pm$0.00 & \textbf{13.35$\pm$3.37} &  & 10.00$\pm$0.00 & 10.55 $\pm$0.44\\\bottomrule
\end{tabular}}
\caption{Comparison of Multi-Krum, Trimmed-Mean, Reputation, and their integration with \textsc{FedQV} under SOTA attacks. The best results are in bold. The results under targeted attacks are 
“ACC / ASR".}
\label{tab:defenses}
\vspace{-10pt}
\end{table}

\subsection{Adaptive Budget}\label{subsec:adaptivebudget}
\vspace{-6pt}
The performance of \textsc{FedQV} with an adaptive budget for the three evaluated methods during the two severe attacks with an increase in the percentage of attackers to 50\% is shown in Figure~\ref{fig:fedqv_rep}. It demonstrates that the combination of \textsc{FedQV} and the reputation model considerably strengthens resistance against Gaussian and Scaling attacks by at least a factor of 26\%. Setting this observation as the alternative hypothesis $H_{1}$ and using the Wilcoxon signed-rank test,  we can reject the null hypothesis $H_{0}$  at a confidence level of 1\% in favour of $H_{1}$.

\subsection{Integration with Byzantine-robust Aggregation}\label{subsec:krum}
\vspace{-6pt}
Our objective is not to position FedQV in competition with existing defence techniques but rather to demonstrate that FedQV can act as a complementary approach to advanced defences. 
\textsc{FedQV} can be seamlessly integrated into Byzantine-robust defence by adapting the weight calculation process.  
We illustrate this with examples using  Muilt-Krum~\cite{blanchard2017machine}, Trim-mean~\cite{yin2018byzantine} and Reputation~\cite{chu2022securing}. 
Figure~\ref{fig:Krum} shows that the accuracy of Multi-Krum increases considerably, especially for local poisoning attacks. 
Table~\ref{tab:multi_krum_targeted} demonstrates 
that the integration of the Multi-Krum, Trimmed-Mean and Reputation method with \textsc{FedQV} leads to superior performance(Higher ACC and lower ASR) compared to the standalone versions.
These findings support that \textsc{FedQV} holds promise as a valuable complementary method to existing defence mechanisms.

\section{Conclusion}
In this paper, we have proposed \textsc{FedQV}, a novel aggregation scheme for FL based on quadratic voting instead of \emph{1p1v}, which is the underlying principle that makes the currently employed \textsc{FedAvg} vulnerable to poisoning attacks. 
The proposed method aggregates global models based upon the votes from a truthful mechanism employed in \textsc{FedQV}. The efficiency of the proposed method has been comprehensively analysed from both a theoretical and an experimental point of view. Collectively, our performance evaluation has shown that \textsc{FedQV} achieves superior performance than \textsc{FedAvg} in defending against various poisoning
attacks. Moreover, \textsc{FedQV} is a reusable module that can be used in conjunction with reputation models to assign unequal voting budgets, other Byzantine-robust techniques, and privacy-preserving mechanisms to provide resistance to both poisoning and privacy attacks. These findings position \textsc{FedQV} as a promising complement to existing aggregation in FL.

\bibliographystyle{iclr2024_conference}
\bibliography{iclr2024_conference}

\clearpage
\appendix
\label{sec:appendix}
We present the related supplements in the following sections. It contains the proof of theoretical analysis section for  Theorem~\ref{th:theorem1}, Theorem~\ref{th:theorem_malicious} and  Theorem~\ref{th:theorem2}, experimental details, and extra results.

\section{Proof of Theoretical Analysis}
\label{sec:th_proof}

\subsection{Assumptions}
\begin{assumption}\label{as:assumption1}
The loss functions  are $L$-smooth, which means they are continuously differentiable and their gradients are Lipschitz-continuous with Lipschitz constant $L>0$, whereas: 
\begin{gather*}
\forall i \in N,\, \forall \bm{w}_{1},\bm{w}_{2} \in \mathds{R}^{d},\;
\| \nabla{\mathcal{L}(\bm{w}_{1}))}-\nabla{\mathcal{L}(\bm{w}_{2}))} \|_{2} \leq L \left \|\bm{w}_{1}-\bm{w}_{2}\right \|_{2}\\
\left \| \nabla{\ell(\bm{w}_1;\mathcal{D})}-\nabla{\ell(\bm{w}_2;\mathcal{D})}\right \|_{2} \leq L \| \bm{w}_{1}-\bm{w}_{2} \|_{2}
\end{gather*}
\end{assumption}

\begin{assumption}\label{as:assumption2} 
The loss function $\ell(\bm{w}_{i},D)$ are $\mu$-strongly convex: 
\begin{gather*}
 \exists \mu > 0,  \forall \bm{w}_{1},\bm{w}_{2} \in \mathds{R}^{d}, \nabla{\ell(\bm{w}^{*};\mathcal{D})} = 0, \nabla{\mathcal{L}(\bm{w}^{*})} = 0\\
     2\left ( \mathcal L(\bm{w}_{1})- \mathcal L(\bm{w}_{2})\right ) \geq 2\left \langle \nabla{\mathcal L(\bm{w}_{2})},  \bm{w}_{1}-\bm{w}_{2}\right \rangle + \mu\left \| \bm{w}_{1}-\bm{w}_{2} \right \|_{2}^{2}\\
 \begin{aligned}    
     2\left ( \ell(\bm{w}_1;\mathcal{D})- l(\bm{w}_2;\mathcal{D})\right )
      \geq  2 \left \langle \nabla{\ell(\bm{w}_2;\mathcal{D})}, \bm{w}_{1}-\bm{w}_{2}\right \rangle  + \mu\left \| \bm{w}_{1}-\bm{w}_{2} \right \|_{2}^{2}
\end{aligned}
\end{gather*}
\end{assumption}

\begin{assumption}\label{as:assumption3}
The expected square norm of gradients $\bm{w}$ is bounded: 
\begin{gather*}
    \forall \bm{w} \in \mathds{R}^{d},  \exists\mathcal{G}_{\bm{w}} < \infty, \mathbb{E}\left \| \nabla{\ell}(\bm{w};\mathcal{D}) \right \|_{2}^{2}\leq \mathcal{G}_{\bm{w}}^{2}
\end{gather*}
\end{assumption}
\begin{assumption}\label{as:assumption4}
The variance of gradients $\bm{w}$ is bounded: 
\begin{gather*}
\forall \bm{w} \in \mathds{R}^{d},  \exists\mathcal{V}_{\bm{w}} < \infty, \mathbb{E}\left \| \nabla{\ell}(\bm{w};\mathcal{D})-\mathbb{E}(\nabla{\ell}(\bm{w};\mathcal{D}) \right \|_{2} ^2 \leq \mathcal{V}_{\bm{w}}  
\end{gather*}
\end{assumption}

\subsection{Proof of Theorem~\ref{th:theorem1} and Theorem~\ref{th:theorem_malicious}}
\subsubsection{Lemmas}~\label{ap:lemmas}
The lemmas we utilize in the proof of Theorem~\ref{th:theorem1} and Theorem~\ref{th:theorem_malicious}, are presented here due to the page limit.
\begin{lemma}
\label{le:lemma4}
Assume Assumption~\ref{as:assumption4} holds, according to our Algorithm~\ref{al:aggregtaion}, it follows that 
\begin{align*}
    \EX\left \|\mathcal{F}(\bm{w}^{t-1}) - \nabla\mathcal{L}(\bm{w}^{t-1})\right \|_{2}^{2} \leq \left(1-2\theta\right)\mathcal{C}\mathcal{V}_{\bm{w}}\sqrt{B}
\end{align*}
Where
\begin{align*}
    \mathcal{F}(\bm{w}^{t-1}) = \sum_{i \in \mathcal{S}^{t-1}} p_{i}^{t-1} \nabla \ell(\bm{w}_{i}^{t-1};\mathcal{D}_{i}^{t-1})
\end{align*}
\end{lemma}

\begin{lemma}
 \label{le:lemma1}
 From Assumption \ref{as:assumption1} and \ref{as:assumption2}, $\mathcal L(\bm{w})$ is $L$-smooth and $\mu$-strongly convex. Then $\forall \bm{w}_{1}, \bm{w}_{2} \in \mathds{R}^{d}$, one has
\begin{align*}
   \langle \nabla \ell(\bm{w}_{1})- \nabla \ell(\bm{w}_{2})&, \bm{w}_{1}  - \bm{w}_{2} \rangle  \geq  \frac{L\mu}{L+\mu}\left \|\bm{w}_{1}-\bm{w}_{2}\right \|_{2}^{2}
    +\frac{1}{L+\mu}\left \| \nabla \ell(\bm{w}_{1}) - \nabla \ell(\bm{w}_{2})\right \|_{2}^{2}
\end{align*}
\end{lemma}

\begin{lemma}
 \label{le:lemma2}
 Assume Assumption~\ref{as:assumption1}, Assumption~\ref{as:assumption2} and Lemma~\ref{le:lemma1} hold, we have
 \begin{align}
      \left \| \bm{w}^{t-1} -  r\nabla\mathcal{L}(\bm{w}^{t-1}) - \bm{w}^{*} \right \|_{2}^{2} & \leq \sum_{i=1}^{N}p_{i}^{t-1}\left \| \bm{w}^{t-1} - \bm{w}^{t-1}_{i}\right \|_{2}^{2}\notag\\
      & + \left(r^{2}\left(1 + L^{2} \right)- \frac{2rL\mu + 1}{L+\mu} \right)\left \| \bm{w}^{t-1} -\bm{w}^{*}\right \|_{2}^{2} 
\end{align}
\end{lemma}

\begin{lemma}
\label{le:lemma3}
Assume Assumption~\ref{as:assumption3} holds, it follows that 
\begin{align*}
      \EX{\sum_{i=1}^{N}p_{i}^{t-1}\left \| \bm{w}^{t-1} - \bm{w}^{t-1}_{i}\right \|_{2}^{2}} \leq \left(E-1\right)^{2}r^{2} \mathcal{G}_{\bm{w}}^{2} 
\end{align*}
\end{lemma}

\subsubsection{Proof of Lemmas}~\label{ap:proof_of_lemmas}
Lemmas~\ref{le:lemma4}, Lemmas~\ref{le:lemma3}, Lemmas~\ref{le:lemma1} and Lemmas~\ref{le:lemma2} are all the lemmas we utilise during the proof of Theorem~\ref{th:theorem1}, and we prove them in that order. Notice, Lemmas~\ref{le:lemma1} are used in the proof Lemmas~\ref{le:lemma2}, and Theorem~\ref{th:theorem1} is proved using Lemmas~\ref{le:lemma4}, Lemmas~\ref{le:lemma3} and Lemmas~\ref{le:lemma2}.

\paragraph{Proof of Lemma \ref{le:lemma4}}
\begin{proof}
Due to Assumption~\ref{as:assumption4} and Algorithm~\ref{al:aggregtaion}, we have
\begin{align}
   \EX\left \|\mathcal{F}(\bm{w}^{t-1}) - \nabla\mathcal{L}(\bm{w}^{t-1})\right \|_{2}^{2} 
   & = \mathrm{Var}\left( \mathcal{F}(\bm{w}^{t-1})\right) 
  = \EX_{\mathcal{S}^{t-1}}\left \|{\sum_{i \in \mathcal{S}^{t-1}}p_{i}^{t-1}}\left( \nabla \ell(\bm{w}_{i}^{t-1};\mathcal{D}_{i}^{t-1}) - \nabla \ell(\bm{w}_{i}^{t-1}) \right) \right \|_{2}^{2} \notag\\
  & = \sum_{i \in \mathcal{S}^{t-1}}\left(p_{i}^{t-1}\right)^{2}\EX\left \|\nabla \ell(\bm{w}_{i}^{t-1};\mathcal{D}_{i}^{t-1}) - \nabla \ell(\bm{w}_{i}^{t-1}) \right \|_{2}^{2} \notag\\
  & \leq \sum_{i \in \mathcal{S}^{t-1}}\left(p_{i}^{t-1}\right)^{2}\mathcal{V}_{\bm{w}} \leq \mathcal{V}_{\bm{w}}\sum_{i \in \mathcal{S}^{t-1}}\left(\frac{v_{i}^{t-1}}{\sum_{i \in \mathcal{S}^{t-1}}v_{i}^{t-1}}\right)^{2} \notag\\
  & \leq \mathcal{V}_{\bm{w}}\frac{\sum_{i \in \mathcal{S}^{t-1}}\left(v_{i}^{t-1}\right)^{2}}{\left(\sum_{i \in \mathcal{S}^{t-1}}v_{i}^{t-1}\right)^{2}}\leq \mathcal{V}_{\bm{w}}\frac{\sum_{i \in \mathcal{S}^{t-1}}\left(v_{i}^{t-1}\right)^{2}}{\sum_{i \in \mathcal{S}^{t-1}}v_{i}^{t-1}}\notag\\
  & \leq \mathcal{V}_{\bm{w}}\sum_{i \in \mathcal{S}^{t-1}}v_{i}^{t-1} 
  \leq \left(1-2\theta\right)qN\mathcal{V}_{\bm{w}}\sqrt{B}
\end{align}
\end{proof}

\paragraph{Proof of Lemma \ref{le:lemma1}}
\begin{proof}
Let $g(\bm{w}) = \ell(\bm{w})-\frac{\varsigma}{2}\left \| \bm{w} \right \|_{2}^{2}$. Base on the Assumption~\ref{as:assumption2}, we have $g(\bm{w})$ is $(L-\varsigma)$-strongly convex.
from \cite{bubeck2015convex} Equation 3.6, we have

\begin{align}
    \langle \nabla \ell(\bm{w}_{1})- \nabla \ell(\bm{w}_{2}),& \bm{w}_{1}  - \bm{w}_{2} \rangle  \geq  \frac{1}{L}\left \| \nabla \ell(\bm{w}_{1}) - \nabla \ell(\bm{w}_{2})\right \|_{2}^{2}
\end{align}

\noindent Hence,

\begin{align}
     \langle \nabla g(\bm{w}_{1})- \nabla g(\bm{w}_{2}), \bm{w}_{1}  - \bm{w}_{2} \rangle  \geq \frac{1}{L-\varsigma}\left \| \nabla g(\bm{w}_{1}) - \nabla g(\bm{w}_{2})\right \|_{2}^{2}
\end{align}

\noindent Now We have

\begin{align}
     \langle \nabla \left ( \ell(\bm{w}_{1})- \frac{\varsigma}{2}\left \| \bm{w}_{1} \right \|_{2}^{2}\right ) & -\nabla \left ( \ell(\bm{w}_{2})- \frac{\varsigma}{2}\left \| \bm{w}_{2} \right \|_{2}^{2}\right ) , \bm{w}_{1}  - \bm{w}_{2} \rangle\notag\\ 
     &  \geq \frac{1}{L+\mu}\left \| \nabla \left ( \ell(\mathbf{\bm{w}_{1}})- \frac{\varsigma}{2}\left \| \bm{w}_{1} \right \|_{2}^{2}\right ) -\nabla \left ( \ell(\bm{w}_{2})- \frac{\varsigma}{2}\left \| \bm{w}_{2} \right \|_{2}^{2}\right )\right \|_{2}^{2}
\end{align}

\noindent And therefore

\begin{align}
     \langle \nabla \ell(\bm{w}_{1})- \nabla \ell(\bm{w}_{2}), \bm{w}_{1} - \bm{w}_{2} \rangle & -  \langle \varsigma \bm{w}_{1}- \varsigma \bm{w}_{2}, \bm{w}_{1} - \bm{w}_{2} \rangle \notag\\ 
     &  \geq \frac{1}{L-\varsigma}\left \| \left ( \nabla \ell(\bm{w}_{1})- \nabla \ell(\bm{w}_{2})  \right ) - \left ( \varsigma \bm{w}_{1}- \varsigma \bm{w}_{2} \right )\right \|_{2}^{2}
\end{align}

\noindent Refer to Assumption \ref{as:assumption1}, we obtain

\begin{align}
    \langle \nabla \ell(\bm{w}_{1})- \nabla \ell(\bm{w}_{2}), \bm{w}_{1}  - \bm{w}_{2} \rangle  & \geq  \frac{L\varsigma}{L-\varsigma}\left \|\bm{w}_{1}-\bm{w}_{2}\right \|_{2}^{2} - \frac{2\varsigma}{L-\varsigma}\left \langle \nabla \ell(\bm{w}_{1}) - \nabla \ell(\bm{w}_{2}), \bm{w}_{1}-\bm{w}_{2}  \right \rangle \notag\\
    & +\frac{1}{L-\varsigma}\left \| \nabla \ell(\bm{w}_{1}) - \nabla \ell(\bm{w}_{2})\right \|_{2}^{2} \notag\\
    & \geq  -\frac{L\varsigma}{L-\varsigma}\left \|\bm{w}_{1}-\bm{w}_{2}\right \|_{2}^{2} +\frac{1}{L-\varsigma}\left \| \nabla \ell(\bm{w}_{1}) - \nabla \ell(\bm{w}_{2})\right \|_{2}^{2} 
\end{align}

\noindent Let $\varsigma = -\mu$, then we conclude the proof of Lemma \ref{le:lemma1}.

\end{proof}

\paragraph{Proof of Lemma \ref{le:lemma2}}
\begin{proof}
\noindent We have 
\begin{align}
\label{eq:proof_A}
    \left \| \bm{w}^{t-1} -  r_{t-1}\nabla\mathcal{L}(\bm{w}^{t-1}) - \bm{w}^{*} \right \|_{2}^{2}  =  \left \| \bm{w}^{t-1} - \bm{w}^{*} \right \|_{2}^{2} 
    \underset{\mathbf{A1}}{\underbrace{- 2r_{t-1}\left \langle \nabla \mathcal L(\bm{w}^{t-1}),\bm{w}^{t-1} - \bm{w}^{*} \right \rangle}} + \underset{\mathbf{A2}}{\underbrace{r_{t-1}^{2} \left \| \nabla\mathcal{L}(\bm{w}^{t-1}) \right \|_{2}^{2}}}
\end{align}

\noindent For part $\mathbf{A1}$ under the Assumption~\ref{as:assumption2}, Lemma~\ref{le:lemma1} and Maclaurin inequality, we have
\begin{align*}
     \mathbf{A1} & = -2r_{t-1}\sum_{i=1}^{N}p_{i}^{t-1} \left \langle \nabla \ell(\bm{w}^{t-1}_{i}),\bm{w}^{t-1} - \bm{w}^{*} \right \rangle \notag\\
    & =-2r_{t-1} \sum_{i=1}^{N}p_{i}^{t-1} \left (\left \langle \nabla \ell(\bm{w}^{t-1}_{i}), \bm{w}^{t-1} - \bm{w}^{t-1}_{i} \right \rangle \right )\notag\\
     & -2r_{t-1} \sum_{i=1}^{N}p_{i}^{t-1} \left (\left \langle \nabla \ell(\bm{w}^{t-1}_{i}), \bm{w}^{t-1}_{i} - \bm{w}^{*} \right \rangle \right )\notag\\
    & \leq \sum_{i=1}^{N}p_{i}^{t-1} \left (r_{t-1}^{2} \left \| \nabla \ell( \bm{w}^{t-1}_{i})\right \|_{2}^{2} + \left \| \bm{w}^{t-1} - \bm{w}^{t-1}_{i}\right \|_{2}^{2} \right) -  \notag\\
    & 2r_{t-1}\sum_{i=1}^{N}p_{i}^{t-1} \left (\frac{1}{L+\mu} \left \| \nabla \ell( \bm{w}^{t-1}_{i})\right \|_{2}^{2} + \frac{L\mu}{L+\mu} \left \| \bm{w}^{t-1}_{i} - \bm{w}^{*}\right \|_{2}^{2} \right) \notag\\
    & = \left(r_{t-1}^{2}-\frac{1}{L+\mu} \right)\sum_{i=1}^{N}p_{i}^{t-1} \left (\left \| \nabla \ell( \bm{w}^{t-1}_{i})\right \|_{2}^{2}  \right) \notag\\
    & + \sum_{i=1}^{N}p_{i}^{t-1}\left \| \bm{w}^{t-1} - \bm{w}^{t-1}_{i}\right \|_{2}^{2}
    -  \frac{2r_{t-1}L\mu}{L+\mu}\left \| \bm{w}^{t-1} -\bm{w}^{*}\right \|_{2}^{2}
\end{align*}

\noindent From Assumption~\ref{as:assumption1} and Jensen inequality, we can derive:

\begin{align}
\label{eq: boundary of gradients of loss function}
    \left \| \nabla \ell( \bm{w}^{t-1}_{i}) - \nabla \ell( \bm{w}^{*})\right \|_{2}^{2} \leq L^2 \left \| \bm{w}^{t-1}_{i} - \bm{w}^{*} \right \|_{2}^{2}
\end{align}

\noindent Hence for $\mathbf{A1}$, by Jensen inequality and Equation~\ref{eq: boundary of gradients of loss function}, we have
\begin{align*}
    \mathbf{A1} & \leq \left(r_{t-1}^{2}-\frac{1}{L+\mu} \right)\sum_{i=1}^{N}p_{i}^{t-1} \left (\left \| \nabla \ell( \bm{w}^{t-1}_{i})\right \|_{2}^{2}  \right) \notag\\
    & + \sum_{i=1}^{N}p_{i}^{t-1}\left \| \bm{w}^{t-1} - \bm{w}^{t-1}_{i}\right \|_{2}^{2}
    -  \frac{2r_{t-1}L\mu}{L+\mu}\left \| \bm{w}^{t-1} - \bm{w}^{*}\right \|_{2}^{2} \notag \\
    & \leq  \left(r_{t-1}^{2}-\frac{1}{L+\mu} \right)\sum_{i=1}^{N}p_{i}^{t-1} \left \| \bm{w}^{t-1}_{i} - \bm{w}^{*} \right \|_{2}^{2} \notag\\
    & + \sum_{i=1}^{N}p_{i}^{t-1}\left \| \bm{w}^{t-1} - \bm{w}^{t-1}_{i}\right \|_{2}^{2}
    -  \frac{2r_{t-1}L\mu}{L+\mu}\left \| \bm{w}^{t-1} -\bm{w}^{*}\right \|_{2}^{2} \notag \\
    & \leq \left(r_{t-1}^{2}- \frac{2r_{t-1}L\mu + 1}{L+\mu} \right)\left \| \bm{w}^{t-1} -\bm{w}^{*}\right \|_{2}^{2}  \notag \\
    & + \sum_{i=1}^{N}p_{i}^{t-1}\left \| \bm{w}^{t-1} - \bm{w}^{t-1}_{i}\right \|_{2}^{2}
\end{align*}

\noindent Similar for $\mathbf{A2}$, we have

\begin{align*}
 \mathbf{A2} & =  r_{t-1}^{2} \left \| \sum_{i=1}^{N}p_{i}^{t-1}  \nabla \ell( \bm{w}^{t-1}_{i})\right \|_{2}^{2} \leq r_{t-1}^{2}  \sum_{i=1}^{N}p_{i}^{t-1} \left \| \nabla \ell( \bm{w}^{t-1}_{i}) \right \|_{2}^{2} \notag\\
 & \leq r_{t-1}^{2}L^{2} \sum_{i=1}^{N}p_{i}^{t-1} \left \| \bm{w}^{t-1}_{i} - \bm{w}^{*} \right \|_{2}^{2}\notag\\
 & = r_{t-1}^{2}L^{2}\left \| \bm{w}^{t-1} -\bm{w}^{*}\right \|_{2}^{2}
\end{align*}

\noindent Then we combine results of $\mathbf{A1}$ and $\mathbf{A2}$ for Equation~\ref{eq:proof_A}, it follows that 

\begin{align}
     \left \| \bm{w}^{t-1} -  r_{t-1}\nabla\mathcal{L}(\bm{w}^{t-1}) - \bm{w}^{*} \right \|_{2}^{2}
     & \leq \left(r_{t-1}^{2}\left(1 + L^{2} \right)- \frac{2r_{t-1}L\mu + 1}{L+\mu} \right)\left \| \bm{w}^{t-1} -\bm{w}^{*}\right \|_{2}^{2}\notag\\
     & +\sum_{i=1}^{N}p_{i}^{t-1}\left \| \bm{w}^{t-1} - \bm{w}^{t-1}_{i}\right \|_{2}^{2}
\end{align}
\end{proof}

\paragraph{Proof of Lemma \ref{le:lemma3}}
\begin{proof}
For each $E$ step FL necessitates a communication.
As a result, for any $t-1 \geq 0 $, $\exists t^{*} \leq t - 1 $ that $ t- t^{*} \leq E, t^{*} \in T$, accordingly $\forall i, j \in \mathcal{S}^{t^{*}}, \bm{w}_{i}^{t^{*}} = \bm{w}_{j}^{t^{*}} =\bm{w}^{t^{*}}$. Then, based on $\EX \left\|  \mathbf{X} - \EX\mathbf{X}\right\|_{2}^{2} \leq \EX\left\|\mathbf{X}\right\|_{2}^{2} $, Jensen inequality and Assumption~\ref{as:assumption3}, we have 

\begin{align}
      \EX{\sum_{i=1}^{N}p_{i}^{t-1}\left \| \bm{w}^{t-1} - \bm{w}^{t-1}_{i}\right \|_{2}^{2}} 
     & =  \EX_{\mathcal{S}^{t^{*}}}{\sum_{i \in \mathcal{S}^{t^{*}}} p_{i}^{t-1}\left \| \left( \bm{w}^{t-1}_{i}- \bm{w}^{t^{*}}\right) - \left(\bm{w}^{t-1} - \bm{w}^{t^{*}}\right)\right \|_{2}^{2}} \notag\\
     & = \EX_{\mathcal{S}^{t^{*}}}\left[ \EX_{\mathcal{S}^{t^{*}}} \left\| \left ( \bm{w}^{t-1}_{i}- \bm{w}^{t^{*}} \right ) - \EX_{\mathcal{S}^{t^{*}}}\left [ \bm{w}^{t-1}_{i}- \bm{w}^{t^{*}} \right]\right\|_{2}^{2}   \right)\notag\\
     & \leq  \EX_{\mathcal{S}^{t^{*}}}\left[ \EX_{\mathcal{S}^{t^{*}}} \left\| \left ( \bm{w}^{t-1}_{i}- \bm{w}^{t^{*}} \right) \right\|_{2}^{2}   \right]\notag\\
     & = \EX_{\mathcal{S}^{t^{*}}}{\sum_{i \in \mathcal{S}^{t^{*}}} p_{i}^{t-1}\left \| \bm{w}^{t-1}_{i}- \bm{w}^{t^{*}}\right \|_{2}^{2}}\notag\\
     & = \EX_{\mathcal{S}^{t^{*}}}{\sum_{i \in \mathcal{S}^{t^{*}}}p_{i}^{t-1}\left \| \sum_{t=t^{*}}^{t-2}\nabla \ell( \bm{w}^{t-1}_{i}, D^{t-1}_{i})\right \|_{2}^{2}}\notag\\
     & \leq \sum_{i \in \mathcal{S}^{t^{*}}} p_{i}^{t-1}\EX_{\mathcal{S}^{t^{*}}}\left(t-1-t^{*}\right)\sum_{t=t^{*}}^{t-2}r_{t-1}^{2}\left\|\nabla \ell( \bm{w}^{t-1}_{i}, D^{t-1}_{i}) \right\|_{2}^{2}\notag\\
     & \leq \sum_{i \in \mathcal{S}^{t^{*}}} p_{i}^{t-1}\left(E-1\right)\sum_{t=t^{*}}^{t-2}r_{t-1}^{2}\left\|\nabla \ell( \bm{w}^{t-1}_{i}, D^{t-1}_{i}) \right\|_{2}^{2} \notag\\
     & \leq \sum_{i \in \mathcal{S}^{t^{*}}} p_{i}^{t-1}\left(E-1\right)\sum_{t=t^{*}}^{t-2}r_{t-1}^{2} \mathcal{G}_{\bm{w}}^{2}\notag\\
     & \leq \sum_{i \in \mathcal{S}^{t^{*}}} p_{i}^{t-1}\left(E-1\right)^{2}r_{t-1}^{2} \mathcal{G}_{\bm{w}}^{2}\notag\\
     & \leq \left(E-1\right)^{2}r_{t-1}^{2} \mathcal{G}_{\bm{w}}^{2} 
\end{align}
\end{proof}

\subsubsection{Theorem~\ref{th:theorem1}}
\begin{theorem}
\label{th:theorem1}
Under Assumptions \ref{as:assumption1}, \ref{as:assumption2}, \ref{as:assumption3} and~\ref{as:assumption4}, and $m = 0$. Choose $\alpha = \frac{L+\mu}{\mu L}$ and $\beta =  2\frac{(L+1)(L+\mu)}{\mu L}$, then \textsc{FedQV} satisfies
  \begin{align}
     \EX \mathcal{L}(\bm{w}^T) - \mathcal{L}(\bm{w}^*)  
  \leq \frac{L}{2\varphi + T}\left( \varphi\EX\left \|\bm{w}^{0}
      - \bm{w}^{*} \right \|_{2}^{2} + \frac{\alpha^{2}}{2}\Delta \right)
  \end{align}
Where
\begin{gather*}
    \Delta = \left(E-1\right)^{2} \mathcal{G}_{\bm{w}}^{2}
    + \left(1-2\theta\right)\mathcal{C}\mathcal{V}_{\bm{w}}\sqrt{B},\; 
    \varphi = \alpha \left ( L+1  \right),\;
    \bm{w}^{t} = \sum_{i=1}^{N}p_{i}^{t}\bm{w}_{i}^{t},\; p_{i}^{t} = \frac{1}{\mathcal{C}}\mathds{1}_{i \in \mathcal{S}^{t}}
\end{gather*}
\end{theorem}

\subsubsection{Proof of Theorem~\ref{th:theorem1}}
\begin{proof}
In $t$ round, due to $m = 0$, we have:
\begin{align}
\label{eq:final_expectation}
      \left \|\bm{w}^{t} - \bm{w}^{*} \right \|_{2}^{2}  & = \left \| \bm{w}^{t-1}-r_{t-1}\mathcal{M}(\bm{w}^{t-1})- \bm{w}^{*} \right \|_{2}^{2} =  \left \| \bm{w}^{t-1}-r_{t-1}\mathcal{F}(\bm{w}^{t-1})- \bm{w}^{*} \right \|_{2}^{2} \notag\\  
     & = \underset{\mathbf{A}}{\underbrace{\left \| \bm{w}^{t-1} - r_{t-1}\nabla\mathcal{L}(\bm{w}^{t-1}) - \bm{w}^{*} \right \|_{2}^{2}}} + \underset{\mathbf{B}}{\underbrace{r_{t-1}^{2}\left \|\mathcal{F}(\bm{w}^{t-1}) - \nabla\mathcal{L}(\bm{w}^{t-1})\right \|_{2}^{2}}}\notag\\
     & + \underset{\mathbf{C}}{\underbrace{2r_{t-1}\left \langle \bm{w}^{t-1} - r_{t-1}\nabla\mathcal{L}(\bm{w}^{t-1}) - \bm{w}^{*}, \mathcal{F}(\bm{w}^{t-1}) - \nabla\mathcal{L}(\bm{w}^{t-1}) \right \rangle}}
\end{align}
Where
\begin{align*}
\mathcal{M}(\bm{w}^{t-1}) = \sum_{i \in \mathcal{S}^{t-1}} p_{i}^{t-1} \mathcal{M}_{i}(\bm{w}_{i}^{t-1})
\end{align*}

\noindent Note that $\EX\mathbf{C} = 0$. For the expectation of $A$, from Lemma~\ref{le:lemma2} and Lemma~\ref{le:lemma3}, it follows that
\begin{align}
    \EX[\mathbf{A}] & =  \EX\left \| \bm{w}^{t-1} - r_{t-1}\nabla\mathcal{L}(\bm{w}^{t-1}) - \bm{w}^{*} \right \|_{2}^{2} \notag\\ 
   & \leq
    \left(r_{t-1}^{2}\left(1 + L^{2} \right)- \frac{2r_{t-1}L\mu + 1}{L+\mu} \right)\left \| \bm{w}^{t-1} -\bm{w}^{*}\right \|_{2}^{2} \notag\\ 
    & + \left(E-1\right)^{2}r_{t-1}^{2} \mathcal{G}_{\bm{w}}^{2}  \notag\\ 
\end{align}

\noindent We use Lemma~\ref{le:lemma4} to bound $\mathbf{B}$, we have
\begin{align}
   \EX[\mathbf{B}] \leq r_{t-1}^{2}\left(1-2\theta\right)qN\mathcal{V}_{\bm{w}}\sqrt{B}
\end{align}

\noindent Hence, we have
\begin{align}
\EX\left \|\bm{w}^{t} - \bm{w}^{*} \right \|_{2}^{2} & \leq 
 r_{t-1}^{2}\left(1 + L^{2} \right)\EX\left \| \bm{w}^{t-1} -\bm{w}^{*}\right \|_{2}^{2} \notag\\
& - \frac{2r_{t-1}L\mu + 1}{L+\mu}\EX\left \| \bm{w}^{t-1} -\bm{w}^{*}\right \|_{2}^{2} + r_{t-1}^{2}\Delta
\end{align}

where 
\begin{align*}
    \Delta = \left(E-1\right)^{2} \mathcal{G}_{\bm{w}}^{2}
    + \left(1-2\theta\right)qN\mathcal{V}_{\bm{w}}\sqrt{B}
\end{align*}

\noindent For the learning rate $r_{t}$, $ \exists \alpha > \frac{L+\mu}{2\mu L}, \exists \beta > 0 $, such that $r_{t} = \frac{\alpha}{\beta + t} \leq \frac{1}{L+1}$. We use mathematical induction to prove the following statement:
\\
\textbf{Proposition}: $\forall t \in \mathds{N}, \EX\left \|\bm{w}^{t} - \bm{w}^{*} \right \|_{2}^{2} \leq \frac{\gamma}{\beta+t}$, where $\gamma = \max\left\{ \frac{(L+\mu)\alpha^{2}\Delta}{2\alpha\mu L - L - \mu}, \beta\EX\left \|\bm{w}^{0} - \bm{w}^{*} \right \|_{2}^{2}\right\}$.\\
Let $P(t)$ be the statement $\EX\left \|\bm{w}^{t} - \bm{w}^{*} \right \|_{2}^{2} \leq \frac{\gamma}{\beta+t}$, we give a proof by induction on $t$.\\
Base case: The statement $P(0)$ holds for $t=0$:
\begin{align*}
    \EX\left \|\bm{w}^{0} - \bm{w}^{*} \right \|_{2}^{2}\leq \frac{\gamma}{\beta}
\end{align*}

\noindent Inductive step: Assume the induction hypothesis that for a particular $j$, the single case $t=j$ holds, meaning $P(j)$ is true:
\begin{align*}
    \EX\left \|\bm{w}^{j} - \bm{w}^{*} \right \|_{2}^{2} \leq \frac{\gamma}{\beta+j}
\end{align*}

\noindent It follows that: 
\begin{align*}
    \EX\left \|\bm{w}^{j+1} - \bm{w}^{*} \right \|_{2}^{2} & \leq \left(r_{t}^{2}\left(1 + L^{2} \right)- \frac{2r_{t}L\mu + 1}{L+\mu}\right)\EX\left \| \bm{w}^{j} -\bm{w}^{*}\right \|_{2}^{2} + r_{t}^{2}\Delta \notag\\
    & \leq \left(1 - \frac{2L\mu\alpha}{(L+\mu)(\beta + j)} \right)\frac{\gamma}{\beta + j}
     + \left(\frac{\alpha}{\beta + j}\right)^{2}\Delta \notag\\
    & =  \left [\frac{\alpha^{2}\Delta}{(\beta + j)^{2}} - \frac{2\alpha\mu L - L - \mu}{(\beta + j)^2(L+\mu)}\gamma\right ] 
     + \frac{\beta+j-1}{\left(\beta + j \right)^{2}}\gamma \notag\\
    & \leq \frac{\gamma}{\beta+j+1}
\end{align*}

\noindent Therefore, the statement $P(j+1)$ also holds true, establishing the inductive step. Since both the base case and the inductive step have been proved as true, by mathematical induction the statement $P(t)$ holds for $\forall t \in \mathds{N}$.

\noindent We choose $\alpha = \frac{L+\mu}{\mu L}$ and $\beta =  2\frac{(L+1)(L+\mu)}{\mu L}$, and we have
\begin{align*}
    \gamma = &\max\left\{ \frac{(L+\mu)\alpha^{2}\Delta}{2\alpha\mu L - L - \mu}, \beta\EX\left \|\bm{w}^{0} - \bm{w}^{*} \right \|_{2}^{2}\right\} \notag\\
    & \leq \frac{(L+\mu)\alpha^{2}\Delta}{2\alpha\mu L - L - \mu} +  \beta\EX\left \|\bm{w}^{0} - \bm{w}^{*} \right \|_{2}^{2} \notag\\
    & = \alpha^2\Delta + 2(L+1)\alpha\EX\left \|\bm{w}^{0} - \bm{w}^{*} \right \|_{2}^{2}
\end{align*}

\noindent Then based on Assumption~\ref{as:assumption1} and Taylor expansion, we have the quadratic upper-bound of $\mathcal{L}(\cdot)$:
\begin{align*}
    \mathcal{L}(\bm{w}_1)-\mathcal{L}(\bm{w}_2) \leq (\bm{w}_1 - \bm{w}_2)^{T}\nabla \mathcal{L}(\bm{w}_2) + \frac{L}{2}\left\|\bm{w}_1 - \bm{w}_2\right\|_{2}^{2}
\end{align*}

\noindent It follows that
\begin{align*}
     \EX \mathcal{L}(\bm{w}^T) - \mathcal{L}(\bm{w}^*)  & \leq \frac{L}{2}\EX\left\|\bm{w}^T - \bm{w}^*\right\|_{2}^{2} \leq \frac{\gamma L}{2(\beta + T)} \notag\\
    & \leq \frac{L}{2\alpha(L+1) + T}\left(\frac{\alpha^{2}}{2}\Delta +\alpha(L+1)\EX\left \|\bm{w}^{0} - \bm{w}^{*} \right \|_{2}^{2} \right)\notag\\
    & = \frac{L}{2\varphi + T}\left( \varphi\EX\left \|\bm{w}^{0}
      - \bm{w}^{*} \right \|_{2}^{2} + \frac{\alpha^{2}}{2}\Delta \right)
\end{align*}
Where
\begin{gather*}
    \Delta = \left(E-1\right)^{2} \mathcal{G}_{\bm{w}}^{2}
    + \left(1-2\theta\right)\mathcal{C}\mathcal{V}_{\bm{w}}\sqrt{B},\; 
    \varphi = \alpha \left ( L+1  \right),\;
    \bm{w}^{t} = \sum_{i=1}^{N}p_{i}^{t}\bm{w}_{i}^{t},\; p_{i}^{t} = \frac{1}{\mathcal{C}}\mathds{1}_{i \in \mathcal{S}^{t}}
\end{gather*}

\end{proof}

\subsubsection{Proof of Theorem~\ref{th:theorem_malicious}}
\begin{proof}
In the $t$ round, we have:
\begin{align}
      \left \|\bm{w}^{t} - \bm{w}^{*} \right \|_{2}^{2}  & = \left \| \bm{w}^{t-1}-r_{t-1}\mathcal{M}(\bm{w}^{t-1})- \bm{w}^{*} \right \|_{2}^{2}\notag\\
      & =  \left \| \bm{w}^{t-1}-r_{t-1}\mathcal{F}(\bm{w}^{t-1})- \bm{w}^{*} + r_{t-1}\mathcal{F} (\bm{w}^{t-1}) - r_{t-1}\mathcal{M}(\bm{w}^{t-1}) \right \|_{2}^{2} \notag\\ 
     & = \underset{\mathbf{A}}{\underbrace{\left \| \bm{w}^{t-1} - r_{t-1}\mathcal{F}(\bm{w}^{t-1}) - \bm{w}^{*} \right \|_{2}^{2}}} + \underset{\mathbf{B}}{\underbrace{r_{t-1}^{2}\left \|\mathcal{F}(\bm{w}^{t-1}) - \mathcal{M}(\bm{w}^{t-1})\right \|_{2}^{2}}}\notag\\
     & + \underset{\mathbf{C}}{\underbrace{2r_{t-1}\left \langle \bm{w}^{t-1} - r_{t-1}\mathcal{F}(\bm{w}^{t-1}) - \bm{w}^{*}, \mathcal{F}(\bm{w}^{t-1}) - \mathcal{M}(\bm{w}^{t-1}) \right \rangle}}
\end{align}
Where
\begin{align*}
\mathcal{M}(\bm{w}^{t-1}) = \sum_{i \in \mathcal{S}^{t-1}} p_{i}^{t-1} \mathcal{M}_{i}(\bm{w}_{i}^{t-1})
\end{align*}

For the expectation of $A$, from Theorem~\ref{th:theorem1}, it follows that
\begin{align}
    \EX[\mathbf{A}] & \leq \frac{1}{2\varphi + t}\left( 2\varphi\EX\left \|\bm{w}^{0}
      - \bm{w}^{*} \right \|_{2}^{2} + \alpha^{2}\Delta \right)
\end{align}

\noindent For $\mathbf{B}$, we have
\begin{align}
   \EX[\mathbf{B}] & = r_{t-1}^{2}\left \|{\sum_{i \in \mathcal{S}^{t-1}}p_{i}^{t-1}} \nabla \ell(\bm{w}_{i}^{t-1}) - \sum_{i \in \mathcal{S}^{t-1}} p_{i}^{t-1} \mathcal{M}_{i}(\bm{w}_{i}^{t-1}) \right\|_{2}^{2}\notag\\
   & = r_{t-1}^{2}\left \|\sum_{i \in \mathcal{S}^{t-1}}p_{i}^{t-1} \left(\nabla \ell(\bm{w}_{i}^{t-1}) - \mathcal{M}_{i}(\bm{w}_{i}^{t-1})\right) \right\|_{2}^{2}\notag\\
   & \leq r_{t-1}^{2}\left \|\sum_{i \in mN}p_{i}^{t-1} \left(\nabla \ell(\bm{w}_{i}^{t-1}) - \mathcal{M}_{i}(\bm{w}_{i}^{t-1})\right) \right\|_{2}^{2}
\end{align}
Where $m$ is the percentage of the malicious parties.

\noindent Due to Equation~\ref{eq:credit}, we have
\begin{align}
\theta \leq \frac{\left \langle \nabla \ell(\bm{w}_{i}^{t-1}), \mathcal{M}_{i}(\bm{w}_{i}^{t-1}) \right \rangle}{\left \|  \nabla \ell(\bm{w}_{i}^{t-1} \right \|\cdot\left \| \mathcal{M}_{i}(\bm{w}_{i}^{t-1})  \right \|} \leq  1-\theta
\end{align}
\noindent Thus,
\begin{align}
\theta \left \|  \nabla \ell(\bm{w}_{i}^{t-1}) \right \|\left \| \mathcal{M}_{i}(\bm{w}_{i}^{t-1})  \right \| \leq \left \langle \nabla \ell(\bm{w}_{i}^{t-1}), \mathcal{M}_{i}(\bm{w}_{i}^{t-1}) \right \rangle \leq  (1-\theta)\left \|  \nabla \ell(\bm{w}_{i}^{t-1}) \right \|\left \| \mathcal{M}_{i}(\bm{w}_{i}^{t-1})  \right \| 
\end{align}
Due to this, we have
\begin{align}
   & \left \|  \nabla \ell(\bm{w}_{i}^{t-1}) \right \|_{2}^{2} - 2(1-\theta)\left \|  \nabla \ell(\bm{w}_{i}^{t-1} \right \|\left \| \mathcal{M}_{i}(\bm{w}_{i}^{t-1})  \right \| + \left \| \mathcal{M}_{i}(\bm{w}_{i}^{t-1})  \right \|_{2}^{2} \notag\\
   & \leq \left \|  \nabla \ell(\bm{w}_{i}^{t-1}) - \mathcal{M}_{i}(\bm{w}_{i}^{t-1}) \right \|_{2}^{2}\notag\\
  & \leq \left \| \nabla \ell(\bm{w}_{i}^{t-1})\right \|_{2}^{2} - 2\theta \left \|  \nabla \ell(\bm{w}_{i}^{t-1}) \right \|\left \| \mathcal{M}_{i}(\bm{w}_{i}^{t-1})  \right \| + \left \| \mathcal{M}_{i}(\bm{w}_{i}^{t-1})  \right \|_{2}^{2}  
\end{align}
Hence we have
\begin{align}
\label{eq:bound_of_sum}
    &\theta(2-\theta)\left \| \nabla \ell(\bm{w}_{i}^{t-1}) \right \|_{2}^{2} + \left \|  (1-\theta)\left \|\nabla \ell(\bm{w}_{i}^{t-1})\right\| - \left\|\mathcal{M}_{i}(\bm{w}_{i}^{t-1})\right\| \right \|_{2}^{2} \notag\\
    & \leq \left \|  \nabla \ell(\bm{w}_{i}^{t-1}) - \mathcal{M}_{i}(\bm{w}_{i}^{t-1}) \right\|_{2}^{2} \notag\\
   & \leq (1-\theta^{2})\left \| \nabla \ell(\bm{w}_{i}^{t-1}) \right \|_{2}^{2} + \left \|  \theta\left \|\nabla \ell(\bm{w}_{i}^{t-1})\right\| - \left\|\mathcal{M}_{i}(\bm{w}_{i}^{t-1})\right\| \right \|_{2}^{2} 
\end{align}
Hence,
\begin{align}
    \theta(2-\theta)\left \| \nabla \ell(\bm{w}_{i}^{t-1}) \right \|_{2}^{2} \leq \left \|  \nabla \ell(\bm{w}_{i}^{t-1}) - \mathcal{M}_{i}(\bm{w}_{i}^{t-1}) \right \|_{2}^{2}
\end{align}
Due to the Triangle Inequality, we have
\begin{align}
    \sqrt{\theta(2-\theta)}\left \| \nabla \ell(\bm{w}_{i}^{t-1}) \right \| \leq \left \|  \nabla \ell(\bm{w}_{i}^{t-1}) - \mathcal{M}_{i}(\bm{w}_{i}^{t-1}) \right \|\leq \left \|  \nabla \ell(\bm{w}_{i}^{t-1})\right\| + \left\| \mathcal{M}_{i}(\bm{w}_{i}^{t-1}) \right \| 
\end{align}
It follows that:
\begin{align}
    \left( \sqrt{\theta(2-\theta)}-1 \right)\left \| \nabla \ell(\bm{w}_{i}^{t-1}) \right\| \leq \left\| \mathcal{M}_{i}(\bm{w}_{i}^{t-1}) \right \|
\end{align}

By incorporating Equation~\ref{eq:bound_of_sum} and leveraging the AM-GM inequality, we can derive the following expression
\begin{align}
    \left \|  \nabla \ell(\bm{w}_{i}^{t-1}) - \mathcal{M}_{i}(\bm{w}_{i}^{t-1}) \right\|_{2}^{2} 
    & \leq (1-\theta^{2})\left \| \nabla \ell(\bm{w}_{i}^{t-1}) \right \|_{2}^{2} + \left \|  \theta\left \|\nabla \ell(\bm{w}_{i}^{t-1})\right\| - \left\|\mathcal{M}_{i}(\bm{w}_{i}^{t-1})\right\| \right \|_{2}^{2} \notag\\
    & \leq  \left(1-\theta^{2} + \left(1+ \theta + \sqrt{\theta(2-\theta)} \right)^{2}\right) \left \|\nabla \ell(\bm{w}_{i}^{t-1}) \right \|_{2}^{2}\notag\\
    & \leq \left( 4 + 6\theta -\theta^2 \right)\left \|\nabla \ell(\bm{w}_{i}^{t-1}) \right \|_{2}^{2}
\end{align}
Therefore,
\begin{align}
    \EX[\mathbf{B}]  \leq  r_{t-1}^{2}\left \|\sum_{i \in mN}p_{i}^{t-1} \left(\sqrt{4+6\theta-\theta^2}\left \|\nabla \ell(\bm{w}_{i}^{t-1}) \right\| \right) \right\|_{2}^{2} \leq \left( 4 + 6\theta -\theta^2\right)m^{2}N^{2}r_{t-1}^{2}\mathcal{G}_{\bm{w}}^{2}
\end{align}
    
Hence for $\mathbf{C}$, we have
\begin{align}
  \EX[\mathbf{C}] 
  & \leq \frac{2mN\mathcal{G}_{\bm{w}}r_{t-1}^{2}\sqrt{4 + 6\theta -\theta^2}}{2\varphi + t}\left( 2\varphi\EX\left \|\bm{w}^{0} - \bm{w}^{*} \right \|_{2}^{2} + \alpha^{2}\Delta \right)
\end{align}

\noindent Then based on Assumption~\ref{as:assumption1} and Taylor expansion, we have the quadratic upper-bound of $\mathcal{L}(\cdot)$:
\begin{align*}
    \mathcal{L}(\bm{w}_1)-\mathcal{L}(\bm{w}_2) \leq (\bm{w}_1 - \bm{w}_2)^{T}\nabla \mathcal{L}(\bm{w}_2) + \frac{L}{2}\left\|\bm{w}_1 - \bm{w}_2\right\|_{2}^{2}
\end{align*}

\noindent It follows that
\begin{align*}
     \EX \mathcal{L}(\bm{w}^T) - \mathcal{L}(\bm{w}^*) & \leq \frac{L}{2}\EX\left\|\bm{w}^T - \bm{w}^*\right\|_{2}^{2} \notag\\
     & \leq \frac{L + 2Lr_{T-1}\varpi}{2\varphi + T}\left( \varphi\EX\left \|\bm{w}^{0}
      - \bm{w}^{*} \right \|_{2}^{2} + \frac{\alpha^{2}}{2}\Delta \right) 
     + \frac{L\varpi^2}{2}
\end{align*}
Where $\varphi = \alpha \left ( L+1  \right)$, $\varpi = mN\mathcal{G}_{\bm{w}}r_{T-1}\sqrt{4 + 6\theta -\theta^2} $
\end{proof}

\subsection{Proof of Theorem~\ref{th:theorem2}}
\subsubsection{Lemmas}
\label{le:lemma_trustful}
\begin{lemma}
\label{le:monotone}
$f$ is monotone: : $\forall v_{-i}$ and $\forall v^{'}_{i} >  v_{i}$, if $f(v_{i},v_{-i}) \in W_{i}$, then $f(v^{'}_{i},v_{-i}) \in W_{i}$.
\end{lemma} 

\begin{lemma}
\label{le:critical value}
In \textsc{FedQV}, $\forall i,v_{i},v_{-i}$ that $f(v_{i},v_{-i})\in W_{i}$, we have that $p_i(v_{i},v_{-i}) = \Phi_i(v_{-i})$, where $\Phi_i$ is the critical value of a monotone function $f$ on a single parameter domain that $\Phi_i(v_{-i}) = \sup_{v_i:f(v_{i},v_{-i})\notin W_{i}} v_{i}$.
\end{lemma}

\subsubsection{Proof of Lemmas}
\paragraph{Proof of Lemmas~\ref{le:monotone}}
\begin{proof}
$\forall v_{-i}$ and $\forall v^{'}_{i} >  v_{i}$, based on the voting scheme, if the party $i$ who submit $s_{i}$ join the aggregation with $v_{i}$, which means $f(v_{i},v_{-i}) \in W_{i}$, then this party can also submit $\forall s^{'}_{i} <  s_{i}$ that lead to $v^{'}_{i} > v_{i}$, and still join the aggregation. In other words, $f(v^{'}_{i},v_{-i}) \in W_{i}$. Thus, $f$ is monotone.
\end{proof}

\paragraph{Proof of Lemmas~\ref{le:critical value}}
\begin{proof}
The number of parties is $\mathcal{C}$ in each round. In voting scheme that follows Equation~\ref{eq:credit}, the parties whose $s_i \leq \theta$ and $s_i \geq 1-\theta $ pay 0 credits voice. After Equation~\ref{eq:vote}, the parties with 0 credit voice or 0 budget gain 0 vote. Assuming there are the top $k$($k<\mathcal{C}$) parties in ranking whose payments are $c_{j \in k}$ ($c_{j \in k} > 0$). Notice in \textsc{FedQV}, the payment function $p_i(v_{i},v_{-i}) = c_{i} = v_{i}^{2}$.

\noindent$\forall j \in k$, if party $j$ pays $c^{'}_{j} > p_j(v_{j},v_{-j}) = \Phi_{i}(v_{-i}) = \sup_{v_i:f(v_{i},v_{-i})\notin W_{i}} v_{i}$, it will still remain in top $k$ and join the aggregation. On the other hand, if party $j$ pays $c^{'}_{j} < p_j(v_{j},v_{-j}) = \Phi_{i}(v_{-i})$, then it will be replaced by the party $k+1$ in the ranking, and party $j$ will not be able to join the aggregation regardless of whether party $k+1$ joins or not. As a result, in order to participate in the aggregation, the parties need to pay critical value, that is, $\forall i,v_{i},v_{-i}$ that $f(v_{i},v_{-i})\in W_{i}$, we have that $p_i(v_{i},v_{-i}) = \Phi_i(v_{-i})$
\end{proof}

\subsubsection{Proof of Theorem~\ref{th:theorem2}}
\begin{proof}
According to Theorem 9.36~\cite{blumrosen2007algorithmic}: a normalised mechanism on a single parameter domain is incentive compatible(truthful) if and only if:\\
(i) The selection rule is monotone.\\ 
(ii) For every party $i$ participants in the aggregation ($v_i > 0$) pays the critical value $\Phi_i(v_{-i}) = \sup_{v_i:f(v_{i},v_{-i})\notin W_{i}} v_{i}$.\\
The first condition (i) and the second one (ii) are proofed in Lemma~\ref{le:monotone} and Lemma~\ref{le:critical value} respectively. Thus, the proposed scheme \textsc{FedQV} is incentive-compatible (truthful).
\end{proof}

\section{\textsc{FedQV} with Adaptive Budgets Algorithm}~\label{ap:algo_budget}
Here we present a concise elucidation of key components of the Algorithm~\ref{al:adaptive_bugdet} as followings:
\begin{itemize}
    \item  \textbf{IRLS} (Iteratively Reweighted Least Squares): IRLS serves as an optimisation technique employed to solve specific regression problems. Within~\cite{chu2022securing}, IRLS is utilised to compute the Subjective Observations of participating clients based on their parameter's confidence score, which is calculated using the repeated-median regression technique.
    \item \textbf{Subjective Observations}: Positive observations denoted by $P_{i}^{t}$ signify acceptance of an update, while negative observations denoted by $N_{i}^{t}$ indicate rejection. Consequently, positive observations enhance a client's reputation, and negative ones have the opposite effect.
    \item \textbf{Reputation Score Calculation}: The reputation score of a client is determined using a subjective logic model, formulated as follows:
$$ R_{i}^{t}=  \frac{\kappa P_{i}^{t} + Wa}{\kappa P_{i}^{t} + \eta N_{i}^{t} + W} $$  
\end{itemize}

Regarding the integration of the reputation model, our objective is to demonstrate how combining \textsc{FedQV} with the reputation model enables the allocation of unequal budgets, thereby enhancing the robustness of standard \textsc{FedQV}. This integration's adaptability extends beyond a single reputation model, allowing customisation to suit various needs. The example presented in the paper serves to showcase the concept's viability.

\section{Experimental Details and Extra Results}
\label{sec:extra_results}
\subsection{Experimental Details}~\label{sec:setting}
Our simulation experiments are implemented with Pytorch framework~\cite{paszke2017automatic} on the cloud computing platform Google Colaboratory Pro (Colab Pro) with access to Nvidia K80s, T4s, P4s and P100s with 25 GB of Random Access Memory. 
Table~\ref{tab:setting} shows the default setting in our experiments.

\begin{table}[!t]
\centering
    \caption{Default experimental settings}
{
\centering
     \begin{tabular}{@{}lll@{}}
\toprule
Explanation                      & Notation                      & Default Setting    \\ \midrule
Budget                & B                             & 25                 \\
Similarity threshold     & $\theta$                      & 0.1                \\ \midrule
The number of parties            & $N$                           & 100                \\
The fraction of selected parties & $\mathcal{C}$                 & 10                 \\
The number of total steps        & T                             & 500                \\
The number of local epochs       & E                             & 5                  \\
Learning rate                    & $r$                           & 0.01               \\
Local batch size                 &                               & 10                 \\
Loss function                    & $\mathcal{L}(\mathbf{\cdot})$ & Cross-entropy \\
Repeating times & 3 \\
\bottomrule
\end{tabular}
    }
    \label{tab:setting}
\end{table}

\subsection{Overview of \textsc{FedQV}}
\begin{figure}[!b]
    \centering
    \includegraphics[width=0.5\textwidth,keepaspectratio]{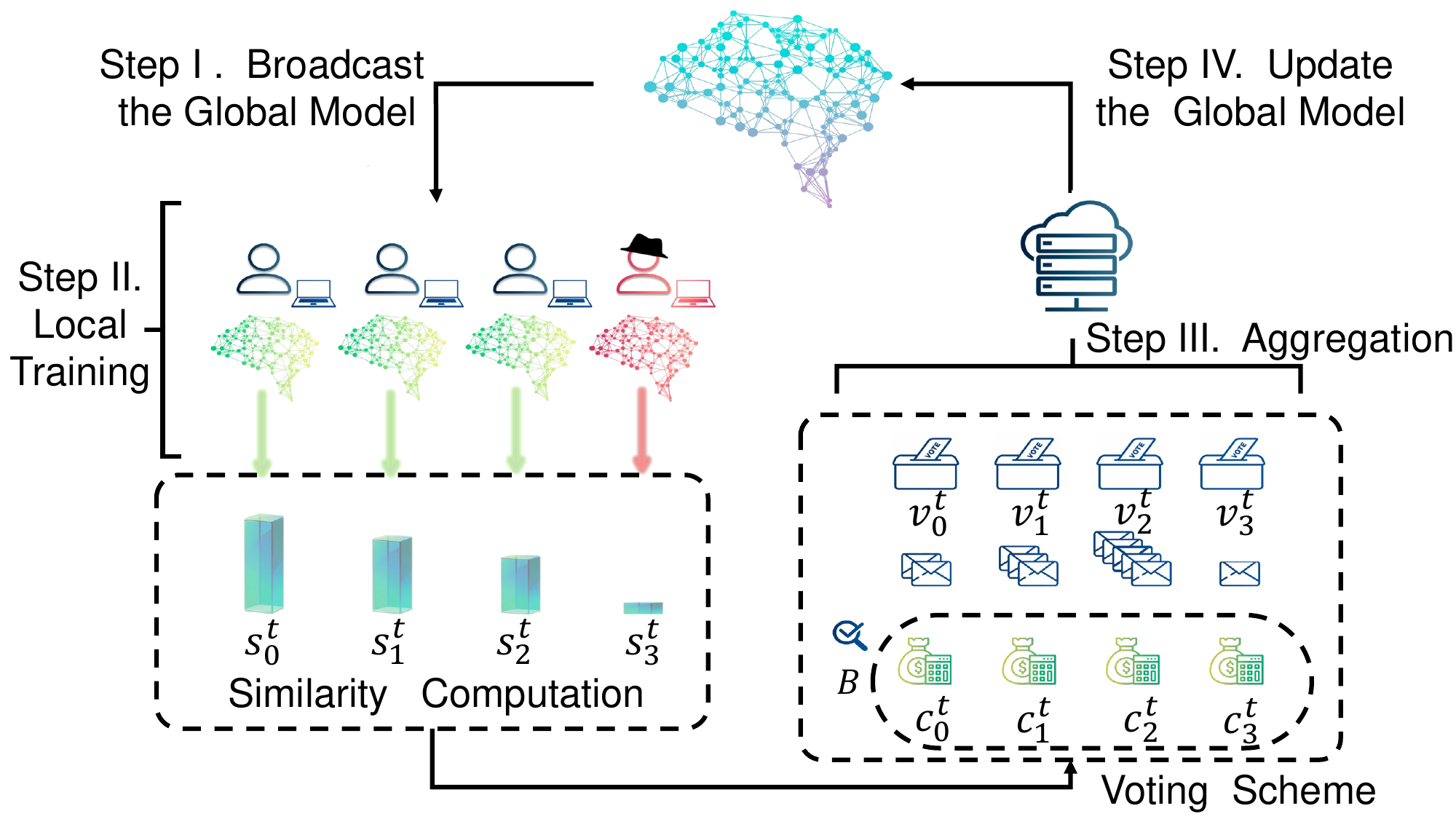}
    \caption{Overview of \textsc{FedQV} algorithm.}
    \label{fig:fedqv}
\end{figure}

Figure~\ref{fig:fedqv} provides an overview of our QV-based aggregation algorithm, which comprises two integral components: "similarity computation" executed on the party side and "voting scheme" managed on the server side. This visual representation encapsulates the essential steps involved in our approach.

\subsection{State-of-the-art Attacks}~\label{ap:attacks}

\noindent \textbf{Labelflip Attack}~\cite{fang2020local}: In the Label-Flip scenario, all the labels of the training data for the malicious clients are set to zero. This scenario simulates a directed attack, with the goal to disproportionally bias the jointly trained model towards one specific class.
This is a data poisoning attack that does not require knowledge of the training data distribution.
Under this attack, the malicious parties  train with clean data but with flipped labels. Specifically, we flip a label $k$ as $K-k-1$, where $K$ is the total class number.

\noindent\textbf{Gaussian Attack}~\cite{zhao2022fedinv}: This attack forges local model updates via Gaussian distribution on the malicious parties. 
malicious parties forge local model updates via Gaussian distribution.

\noindent\textbf{Krum Attack}~\cite{fang2020local}: Malicious parties craft poisoned local model updates opposite from benign ones, and enable them to circumvent the defence of Krum~\cite{blanchard2017machine}.

\noindent\textbf{Trim Attack}~\cite{fang2020local}
The poisoned local model updates constructed by malicious parties are optimised for evading the Trim-mean and Median~\cite{yin2018byzantine}. 

\noindent\textbf{Min-Max Attack}~\cite{shejwalkar2021manipulating}In order to ensure that the malicious gradients closely align with the benign gradients within the clique, attackers strategically compute the malicious gradient. This computation is carried out to limit the maximum distance of the malicious gradient from any other gradient, which is constrained by the maximum distance observed between any two benign gradients.

\noindent\textbf{Min-Sum}~\cite{shejwalkar2021manipulating}
The Min-Sum attack enforces an upper bound on the sum of squared distances between the malicious gradient and all the benign gradients. This upper bound is determined by the sum of squared distances between any one benign gradient and the rest of the benign gradients.

\noindent The targeted poisoning attacks include:

\noindent\textbf{Backdoor Attack}~\cite{gu2019badnets}
Malicious parties inject specific backdoor triggers into the training data and modify their labels to the attacker-chosen target label. Specifically, we use the same backdoor pattern trigger and attacker-chosen target label as in ~\cite{bagdasaryan2021blind} as our trigger and set the attacker-chosen target label as 5.

the backdoor can be introduced into a model by an attacker who poisons the training data with specially crafted inputs. A backdoor transformation applied to any input causes the model to mis-classify it to an attacker-chosen label The pattern must be applied by the attacker during local training, by modifying the digital image.

\noindent\textbf{Scaling attack}~\cite{bagdasaryan2020backdoor}
The malicious parties generate poisoned local model updates by backdoor attack and only launch this attack during the last communication round after scaling these updates by a factor of $N$.

\noindent\textbf{Neurotoxin attack}~\cite{zhang2022neurotoxin}
In this attack, the adversary starts by downloading the gradient from the previous round and employs it to approximate the benign gradient for the upcoming round. The attacker identifies the top-k\% coordinates of the benign gradient and treats them as the constraint set. Over several epochs of Projected Gradient Descent (PGD), the attacker computes gradient updates on the manipulated dataset and projects this gradient onto the constraint set, which consists of the bottom-k\% coordinates of the observed benign gradient. PGD is employed to approach the optimal solution within the span of the bottom-k\% coordinates. We adopt the original parameter setting from the paper, where k is set to 0.1.

\noindent\textbf{QV-Adaptive attack}
We introduce an adaptive attack, \textbf{QV-Adaptive}, tailored for \textsc{FedQV}, utilising the Aggregation-agnostic optimizations~\cite{shejwalkar2021manipulating} within the LMP framework~\cite{fang2020local}.
This attack manipulates both the similarity score and the local model, following the procedure below:

1) The malicious party $i$ generates benign updates $\bm{w}_i^t$ using clean data $\mathcal{D}_i$ in round $t$ and calculates the corresponding similarity score;
 
2) malicious parties (with counts of $m$) collectively normalise all the similarity scores and employ the Aggregation-agnostic Min-Max optimisation to select the optimal similarity score. This optimisation objective aims to increase the likelihood of the score being accepted by the server.

3) the adaptive attack focuses on local model poisoning to optimise the following problem:
    \begin{align}
& \max {\nu} && \\
\text{s.t.} \quad & \bm{w}_{i \in m}^{t'} = \text{FedQV}(\bm{w}_{1}^{t}, \bm{w}_{2}^{t}, \ldots, \bm{w}_{m}^{t}) && \\
& \bm{w}_{i \in m}^{t'} = \bm{w}_i^t - \nu \hat{d} &&
\end{align}
    
Here, $\hat{d}$ represents a column vector encompassing the estimated changing directions of all global model parameters. The variables $\bm{w}_{i \in m}^{t}$
and $\bm{w}_{i \in m}^{t'}$
correspond to the local model before and after the attack. The parameter $\nu$ denotes the extent of the attack's impact on the model.

\subsection{Preliminary Results}~\label{ap:preliminary}
In \textsc{FedAvg}, for example, if the malicious parties hold a substantial amount of local data and poison it, the accuracy of the global model would suffer owing to its aggregation rule. 
We use \textsc{FedQV} to solve this dilemma.

To demonstrate how \textsc{FedQV} constrain the influence of malicious parties, we
consider two benign and one malicious party who conduct backdoor attacks with the amount of training data $\left\{1,1,2\right\}$. We train a multi-layer CNN for 10 rounds in the MNIST dataset same as in Section~\ref{sec:experiments}. The test accuracy is shown in Figure~\ref{fig:toy_model} in which the sides of the triangle correspond to the different parties and the position inside the triangle corresponds to their aggregation weights. 

We observed that compared to \textsc{FedAvg} with the weight  $\left\{ 1,1,2\right\}$, QV, with the weight setup $\left\{ 1,1,\sqrt{2}\right\}$, achieves higher accuracy. This suggests that QV can enhance performance by restraining the influence of attackers within \textsc{FedAvg}. Consequently, when QV is integrated into FL with masked voting rules and a limited budget, as in \textsc{FedQV}, it effectively excludes the malicious party and yields higher accuracy, represented by the weight configuration $\left\{ 1,1,0\right\}$.


To demonstrate how \textsc{FedQV} compute the aggregation weights, consider the following scenario: there are 10 parties in the FL system, and 7 of them are attackers.
The training consists of 10 communication rounds, during which attackers execute backdoor attacks. The rest of the settings are the same as the default. 
\begin{figure}[!]
    \centering
    \includegraphics[width=0.7\columnwidth]{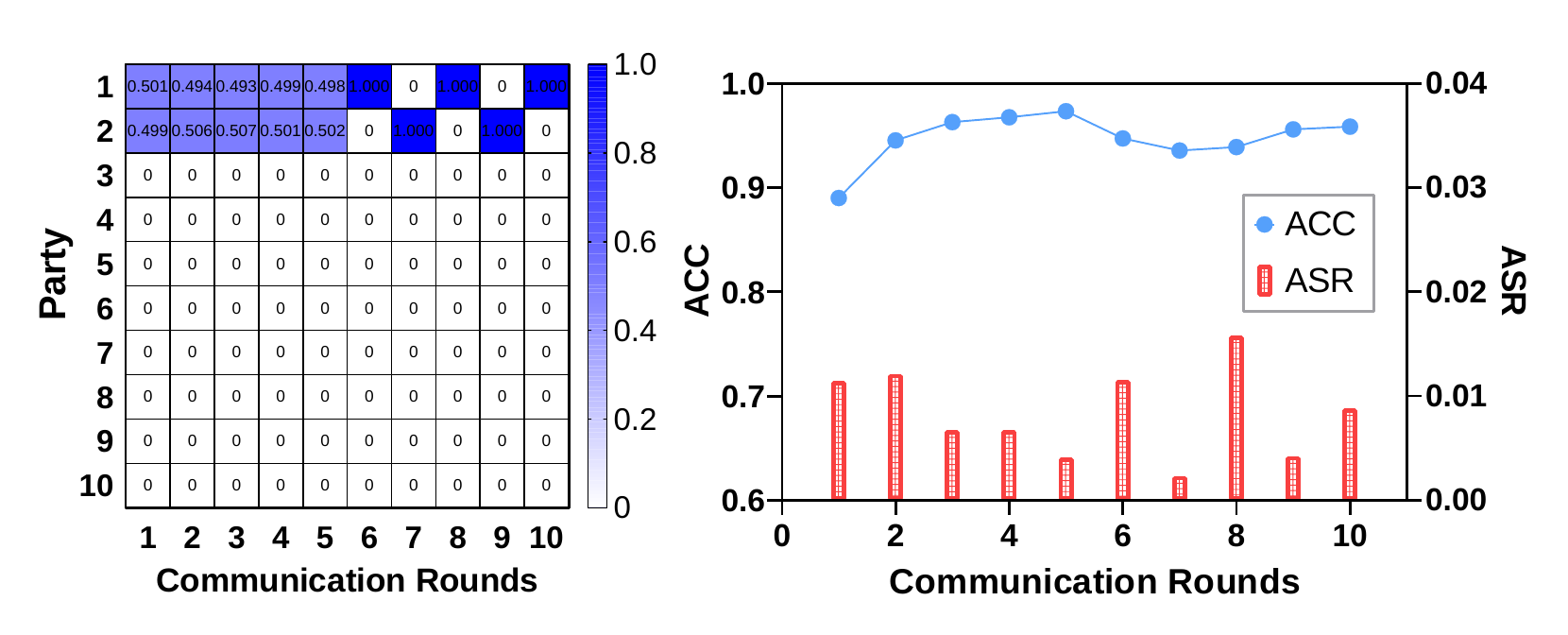}
    \caption{\textsc{FedQV} aggregation weights of each party(left), ACC and ASR for global model(right), for 10 communication rounds in MNIST dataset under Backdoor attack}
    \label{fig:fedqv_weight}
\end{figure}
The result is shown in Figure~\ref{fig:fedqv_weight}.
In the left of Figure~\ref{fig:fedqv_weight}, the first three parties are benign, and the rest are malicious.
We observe that the aggregation weights of malicious parties are 0, implying that \textsc{FedQV} succeed in eliminating their influence. As a result, ASR is quite low, and the accuracy of the global model is unaffected. This demonstrates that even if malicious parties dominate the majority, they do not prevail in damaging the global model. 

\subsection{Non-IID Degree}~\label{ap:non-iid}
To concerning datasets with non-IID data across clients, our experiments incorporate datasets with non-IID characteristics, with a non-IID degree ($\iota$ ) of 0.9. Moreover, we have examined the performance of \textsc{FedQV} and \textsc{FedAvg} across varying levels of non-IID data, spanning from 0.1 to 0.9, as depicted in Table~\ref{tab:non-iid}.

\begin{table}[]
\centering
\begin{tabular}{@{}crlllll@{}}
\toprule
\multicolumn{2}{c}{Non-IID} & \multicolumn{1}{c}{0.1} & \multicolumn{1}{c}{0.3} & \multicolumn{1}{c}{0.5} & \multicolumn{1}{c}{0.7} & \multicolumn{1}{c}{0.9} \\ \midrule
\multirow{2}{*}{FedQV}  & ACC(\%) & 84.94 & 86.01 & 83.88 & 81.37 & 75.96 \\
                        & ASR(\%) & 3.39  & 4.55  & 17.64 & 20.59 & 24.18 \\
\midrule
\multirow{2}{*}{FedAvg} & ACC(\%) & 81.27 & 81.1  & 82.44 & 80.77 & 65.68 \\
                        & ASR(\%) & 3.37  & 13.39 & 20.84 & 22.99 & 60.35 \\ \bottomrule
\end{tabular}
\caption{Comparison of Accuracy (ACC) and Attack Success Rate (ASR) for FedQV and FedAvg under Backdoor Attack over 100 epochs with varying Non-IID Degrees on Fashion-MNIST Dataset.}
\label{tab:non-iid}
\end{table}

These results demonstrate that as the non-IID degree increases among the clients, the performance of the global model declines. Notably,  \textsc{FedQV} consistently maintains a superior performance compared to \textsc{FedAvg}, even when confronted with different degrees of data heterogeneity under attack conditions.

\subsection{Impact of Hyperparameters}
\label{ap:hyper}
As noted, Theorem~\ref{th:theorem_malicious} provides general guidelines for tuning, and the findings from our grid search. As shown in Remark~\ref{re:remark2}, the error rate is influenced by $B$ and $\theta$. To demonstrate the impact of these two hyper-parameters, we grid search $B$ in $\left [ 10,20,30,40,50 \right ]$ and $\theta$ in $\left [ 0.1,0.2,0.3,0.4,0.5 \right ]$. The setup is the same as on the MNIST dataset under the backdoor attack with 30\% malicious parties. 

\begin{figure}[!]
    \centering\includegraphics[width=0.6\columnwidth]{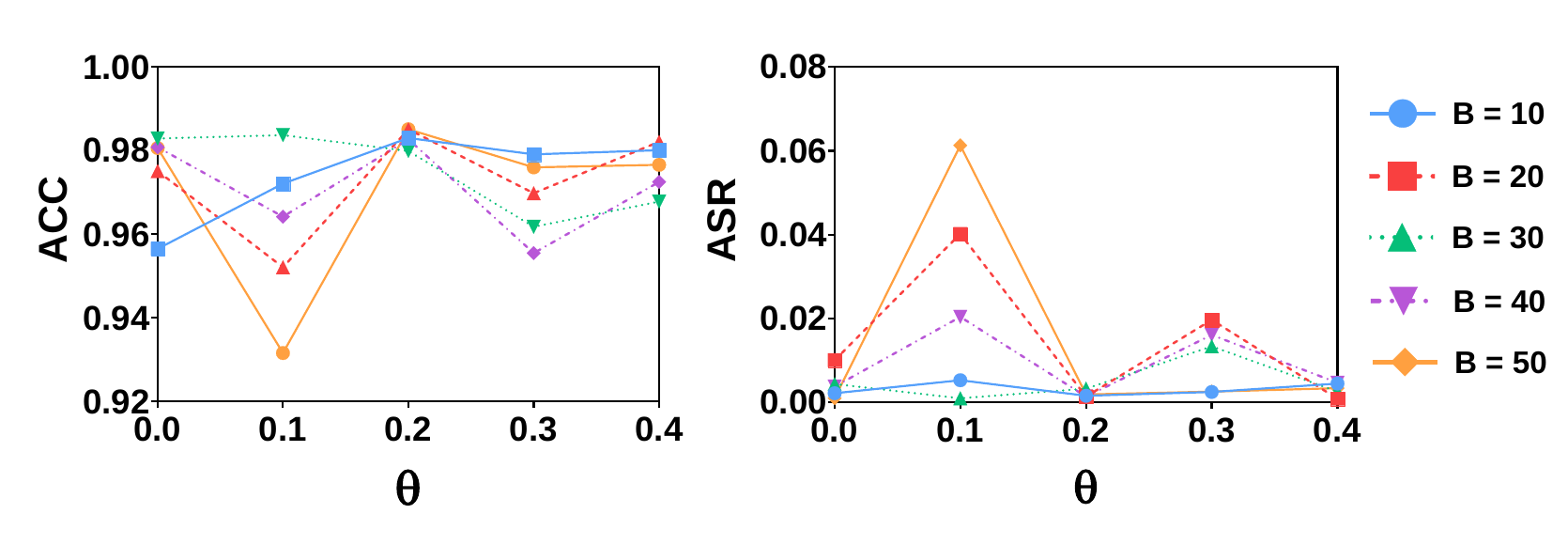}
    \caption{ACC and ASR as we vary the hyperparameters similarity threshold $\theta$ and budget $B$.}
    \label{fig:hyper}
\end{figure}

Figure~\ref{fig:hyper} shows that the optimal values of $B$ and $\theta$ are 30 and 0.2 respectively in this case. As $B$ increases, there is a decline in ACC coupled with an increase in ASR.
These results indicate that \textsc{FedQV}'s performance is not highly sensitive to the hyperparameters, as long as they are chosen in a reasonable range. The approach of combining theoretical guidelines with an exhaustive search to find optimal parameters is a commonly adopted strategy used in similar works.

We can see from Theorem~\ref{th:theorem_malicious}, that the number of malicious devices $m$ will affect the algorithm, and more malicious devices can lead to increased damage. However, this does mean the server needs to know the number of malicious devices to do the fine-tuning. We agree that determining optimal parameters can be challenging, especially in the absence of complete knowledge about the FL system.

A better tuning is possible if more information is available. For specific tasks, more information can indeed be collected from which practical parameter sets can be extracted either via exhaustive search or via simpler online algorithms using trial and error. We will add this to our future work and consider it when we study particular domain-specific problems using our method.

\subsection{Extra Results for Integration with Byzantine-Robust Aggregation}
Table~\ref{tab:multi_krum_targeted} demonstrates that when Multi-Krum
are integrated with FedQV, its ACC increases by at least 28\%, and its ASR decreases by at
least 70\%.
\begin{table}[!]
\centering
\resizebox{0.8\columnwidth}{!}{%
\begin{tabular}{@{}rrrcrrcrrcrrc@{}}\toprule
& \multicolumn{2}{c}{MNIST} & \phantom{abc}& \multicolumn{2}{c}{Fashion-MNIST} &
\\
\cmidrule{2-3} \cmidrule{5-6} 
& Multi-Krum & + \textsc{FedQV} && Multi-Krum & + \textsc{FedQV}  \\ \midrule
Backdoor\\
ACC & 70.20$\pm$9.99 & \textbf{89.96$\pm$1.85}  && 33.24$\pm$13.24 & \textbf{70.89$\pm$3.17} \\
ASR & 32.03$\pm$11.20& \textbf{9.59$\pm$2.28}&& 68.87$\pm$17.77& \textbf{9.72$\pm$4.50} \\
Scaling\\
ACC & 68.35$\pm$16.76 & \textbf{96.55$\pm$0.41}&& 59.43$\pm$14.22 & \textbf{82.48$\pm$0.24}\\
ASR & 33.65$\pm$19.15& \textbf{0.41$\pm$0.06} && 33.64$\pm$19.08 & \textbf{0.91$\pm$0.18} \\
\bottomrule
\end{tabular}}
\vspace{3pt}
\caption{Comparison of Multi-Krum and Multi-Krum + \textsc{FedQV} under targeted attacks with 30\% malicious parties. The best results are in bold.}
\label{tab:multi_krum_targeted}
\vspace{-10pt}
\end{table}

\begin{table}[!]
\begin{tabular}{@{}rcccc@{}}
\toprule
           & Trimmed-Mean & Trimmed-Mean-QV & Trimmed-Mean & Trimmed-Mean-QV \\ \midrule
Neurotoxin & ACC(\%)      & ACC(\%)         & ASR(\%)      & ASR(\%)         \\
1\%        & 86.43        & 86.74           & 0.76         & 0.56            \\
5\%        & 84.96        & 86.34           & 0.92         & 0.72            \\
10\%       & 85.64        & 86.09           & 2.86         & 1.80            \\ \midrule
Backdoor   &              &                 &              &                 \\
1\%        & 84.99        & 85.67           & 0.57         & 0.52            \\
5\%        & 84.83        & 85.66           & 0.93         & 0.46            \\
10\%       & 85.45        & 85.06           & 2.27         & 1.79            \\ \bottomrule
\end{tabular}
\caption{Comparison of Trimmed-Mean and Trimmed-Mean Integrated with FedQV Methods under Targeted Attacks (Backdoor and Neurotoxin) Across Varying Percentages of Malicious Parties.}
\label{tab:target_small}
\end{table}

\end{document}